\documentclass[11pt,english]{scrartcl}
\usepackage[T1]{fontenc}
\usepackage[latin9]{inputenc}
\usepackage{geometry}
\usepackage{array}
\usepackage{float}
\usepackage{multirow}
\usepackage{amsmath}
\usepackage{amssymb}
\usepackage{graphicx}
\usepackage{setspace}
\usepackage[authoryear]{natbib}
\makeatletter

\providecommand{\tabularnewline}{\\}

\usepackage{url}
\usepackage{natbib}
\bibpunct{[}{]}{;}{a}{,}{,}
\usepackage[font=small,format=plain,labelfont=bf, margin=0.5cm]{caption}
\usepackage{hyperref}
\hypersetup{colorlinks=flase,citecolor=black,linkcolor=black}
\usepackage{multirow}
\usepackage[english]{babel}
\usepackage{array}
\usepackage{booktabs}
\date{}

\makeatother

\usepackage{babel}
\begin{document}

\title{Variational Data Assimilation via Sparse Regularization}

\author{A.M. Ebtehaj\textsuperscript{1,2}, M. Zupanski\textsuperscript{3},
G. Lerman\textsuperscript{2}, E. Foufoula-Georgiou\textsuperscript{1}\\
{\small \textsuperscript{1}Department of Civil Engineering, Saint
Anthony Falls Laboratory, University of Minnesota}\\
{\small \textsuperscript{2}School of Mathematics, University of Minnesota}\\
{\small \textsuperscript{3}Cooperative Institute for Research in
the Atmosphere, Colorado State University, Fort Collins, Colorado}}
\maketitle
\begin{abstract}
This paper studies the role of sparse regularization in a properly
chosen basis for variational data assimilation (VDA) problems. Specifically,
it focuses on data assimilation of noisy and down-sampled observations
while the state variable of interest exhibits sparsity in the real
or transformed domain. We show that in the presence of sparsity, the
$\ell_{1}$-norm regularization produces more accurate and stable
solutions than the classic data assimilation methods. To motivate
further developments of the proposed methodology, assimilation experiments
are conducted in the wavelet and spectral domain using the linear
advection-diffusion equation.
\end{abstract}

\section{Introduction}

Environmental prediction models are initial value problems and their
forecast skills highly depend on the quality of their initialization.
Data assimilation (DA) seeks the best estimate of the initial condition
of a (numerical) model, given observations and physical constraints
coming from the underlying dynamics \citep[see,][]{[Dal93],[Kal03]}.
This important problem is typically addressed by two major classes
of methodologies, namely sequential and variational methods \citep{[IdeCGL97]}.
The sequential methods are typically built on the theory of mathematical
filtering and recursive weighted least-squares \citep[among others]{[GilCTBI81],[Ghil89],[GhiM91],[Eve94a],[And01],[MorKGS05],[ZhoEM05],[Lee10]},
while the variational methods are mainly rooted in the theories of
constrained mathematical optimization and batch mode weighted least-squares
(WLS) \citep[e.g.,][among others]{[Sas70a],[Lor86],[Lor88],[CouT90],[Zup93]}.

Although, recently the sequential methods have received a great deal
of attention, the variational methods are still central to the operational
weather forecasting systems. Classic formulation of the variational
data assimilation (VDA) typically amounts to defining a (constrained)
weighted least-squares penalty function whose optimal solution is
the best estimate of the initial condition, the so-called \emph{analysis}
state. This penalty function typically encodes the weighted sum of
the costs associated with the distance of the unknown true state to
the available observations and previous model forecast, the so-called
\emph{background} state. Indeed, the penalty function enforces the
solution to be close enough to both observations and background state
in the weighted mean squared sense, while the weights are characterized
by the observations and the background error covariance matrices.
On the other hand, the constraints typically enforce the analysis
to follow the underlying prognostic equations in a weak or strong
sense \citep[see,][p.369]{[Sas70a],[Dal93]}. Typically, when we constrain
the analysis only to the available observations and the background
state at every instant of time, the variational data assimilation
problem is called 3D-Var \citep[e.g.,][]{[Lor86],[ParD92],[Loretal00],Kleist2009}.
On the other hand, when the analysis is also constrained to the underlying
dynamics and available observations in a window of time, the problem
is called 4D-Var \citep[e.g.,][]{[Zup93],[RabETAL00],[Rawetal07]}.

Inspired by the theories of smoothing spline and kriging interpolation
in geostatistics, the first signs of using regularization in variational
data assimilation trace back to the work by \citet{[WahW80]} and
\citet{[Lor86]}, where the motivation was to impose  smoothness over
the class of twice differentiable analysis states. More recently,
\citet{[JohNH05]} argued that, in the classic VDA problem, the sum
of the squared or $\ell_{2}$-norm of the weighted background error
resembles the Tikhonov regularization \citep{[Tik77]}. Specifically,
by the well-known connections between the Tikhonov regularization
and spectral filtering via singular value decomposition (SVD) \citep[e.g., see][]{[Han1998],[GolHO99],[Han06]},
a new insight was provided into the interpretation and the stabilizing
role of the background state on the solution of the classic VDA problem
\citep[see,][]{[JohHN06]}. Instead of using the $\ell_{2}$-norm
of the background error, \citet{[FreNB10]} and \citet{[BudFN11]}
suggested to modify the classic VDA cost function using the sum of
the absolute values or $\ell_{1}$-norm of the weighted background
error. This assumption requires to statistically assume that the background
error is heavy tailed and can be well approximated by the family of
Laplace densities \citep[e.g.,][]{[Tib96],[LewS00]}. For data assimilation
of sharp atmospheric fronts, \citet{[FreNB12]} kept the classic VDA
cost function while further proposed to regularize the analysis state
by constraining the $\ell_{1}$-norm of its derivative coefficients.
\citet{[EbtF12b]} also used Huber-norm regularization to assimilate
noisy and low-resolution observations into the dynamics of the heat
equation.

In this study, we extend the previous studies \citep[e.g.,][]{[FreNB12],[EbtF12b]}
in regularized variational data assimilation (RVDA) by: (a) proposing
a generalized regularization framework for assimilating low-resolution
and noisy observations while the initial state of interest exhibits
sparse representation in an appropriately chosen basis (i.e., wavelet,
discrete cosine transform); (b) demonstrating the promise of the methodology
in an assimilation example using advection-diffusion dynamics with
different error structure; and (c) proposing an efficient solution
method for large-scale data assimilation problems.

The concept of sparsity plays a central role in this paper. By definition,
a state of interest is sparse in a pre-selected basis, if the number
of non-zero elements of its expansion coefficients in that basis (e.g.,
wavelet coefficients) is significantly smaller than the overall dimension
of the state in the observational space. Here, we show that if sparsity
in a pre-selected basis holds, this prior information can serve to
improve the accuracy and stability of data assimilation problems.
To this end, using prototype studies, different initial conditions
are selected, which are sparse under the wavelet and spectral discrete
cosine transformation (DCT). The promise of the $\ell_{1}$-norm RVDA
is demonstrated via assimilating down-sampled and noisy observations
in a 4D-Var setting by strongly constraining the solution to the governing
advection-diffusion equation. In a broader context, we delineate the
roadmap and explain how we may exploit sparsity, while the underlying
dynamics and observation operator might be nonlinear. Particular attention
is given to explain Monte Carlo driven approaches that can incorporate
a sparse prior in the context of ensemble data assimilation.

Section 2 reviews the classic variational data assimilation problem.
In Section 3, we discuss the concept of sparsity and its relationship
with $\ell_{1}$-norm regularization in the context of VDA problems.
Results of the proposed framework and comparisons with classic methods
are presented in Section 4. Section 5, is devoted to conclusions and
ideas for future research, mainly focusing on the use of ensemble-based
approaches to address sparse promoting VDA in nonlinear dynamics.
Algorithmic details and derivations are presented in Appendix A.

\section{Classic Variational Data Assimilation }

At the time of model initialization $t_{0}$, the goal of data assimilation
can be stated as that of obtaining the \emph{analysis }state as the
best estimate of the true initial state, given noisy and low-resolution\emph{
observations} and the erroneous \emph{background }state, while the
analysis needs to consistent with the underlying model dynamics. The
background state in VDA is often considered to be the previous-time
forecast provided by the prognostic model. By solving the VDA problem,
the analysis is then being used as the initial condition of the underlying
model to forecast the next time step and so on. In the following,
we assume that the unknown true state of interest at the initial time
$t_{0}$ is an $m$-element column vector in discrete space denoted
by $\mathbf{x}_{0}=\left[x_{0,1},\,\ldots,\, x_{0,m}\right]^{{\rm T}}\in\mathbb{R}^{m}$,
the noisy and low-resolution observations in the time interval $\left[t_{0},\,\ldots,\, t_{k}\right]$
are $\mathbf{y}_{i}\in\mathbb{R}^{n}$, $i=1,\,\ldots,\, k$ , where
$n\ll m$. Suppose that the observations are related to the true states
by the following observation model
\begin{equation}
\mathbf{y}_{i}=\mathcal{H}\left(\mathbf{x}_{i}\right)+\mathbf{v}_{i},\label{eq:1}
\end{equation}
where $\mathcal{H}:\,\mathbb{R}^{m}\rightarrow\mathbb{R}^{n}$ denotes
the nonlinear observation operator that maps the state space into
the observation space, and $\mathbf{v}_{i}\sim\mathcal{N}\left(0,\,\mathbf{R}_{i}\right)$
is the Gaussian observation error with zero mean and covariance $\mathbf{R}_{i}$.

Taking into account the sequence of available observations, $\mathbf{y}_{i}\in\mathbb{R}^{n}$
, $i=0,\,\ldots k$, and denoting the background state and its error
covariance by $\mathbf{x}_{0}^{b}\in\mathbb{R}^{m}$ and $\mathbf{B}\in\mathbb{R}^{m\times m}$,
the 4D-Var problem amounts to obtaining the analysis at initial time
as the minimizer of the following WLS cost function:
\begin{equation}
\mathcal{J}_{4D}(\mathbf{x}_{0},\,\mathbf{x}_{1},\,\ldots,\,\mathbf{x}_{k})=\sum_{i=0}^{k}\left(\frac{1}{2}\left\Vert \mathbf{y}_{i}-\mathcal{H}\left(\mathbf{x}_{i}\right)\right\Vert _{\mathbf{R}_{i}^{-1}}^{2}\right)+\frac{1}{2}\left\Vert \mathbf{x}_{0}^{b}-\mathbf{x}_{0}\right\Vert _{\mathbf{B}^{-1}}^{2},\label{eq:2}
\end{equation}

while the solution is constrained to the underlying model equation,
\begin{equation}
\mathbf{x}_{i}=\mbox{\ensuremath{\mathcal{M}}}_{0,\, i}(\mathbf{x}_{0}),\,\, i=0,\ldots,k.\label{eq:3}
\end{equation}

Here, $\left\Vert \mathbf{x}\right\Vert _{\mathbf{A}}^{2}=\mathbf{x}^{{\rm T}}\mathbf{A}\mathbf{x}$
denotes the \emph{quadratic-norm,} while $\mathbf{A}$ is a positive
definite matrix and the function $\mathcal{M}_{0,\, i}:\mathbb{R}^{m}\rightarrow\mathbb{R}^{m}$
is a nonlinear model operator that evolves the initial state in time
from $t_{0}$ to $t_{i}$.

Let us define $\mathbf{M}_{0,\, i}$ to be the Jacobian of $\mathcal{M}_{0,\, i}$
and restrict our consideration only to a linear observation operator,
that is $\mathcal{H}\left(\mathbf{x}_{i}\right)=\mathbf{H}\mathbf{x}_{i}$,
and thus the 4D-Var cost function reduces to
\begin{equation}
\mathcal{J}_{4D}(\mathbf{x}_{0})=\sum_{i=0}^{k}\left(\frac{1}{2}\left\Vert \mathbf{y}_{i}-\mathbf{H}\mathbf{M}_{0,\, i}\,\mathbf{x}_{0}\right\Vert _{\mathbf{R}_{i}^{-1}}^{2}\right)+\frac{1}{2}\left\Vert \mathbf{x}_{0}^{b}-\mathbf{x}_{0}\right\Vert _{\mathbf{B}^{-1}}^{2}.\label{eq:4}
\end{equation}

By defining $\underline{\mathbf{y}}=\left[\mathbf{y}_{0}^{{\rm T}},\,\ldots,\,\mathbf{y}_{k}^{{\rm T}}\right]^{{\rm T}}\in\mathbb{R}^{N}$,
where $N=n(k+1)$, $\mathbf{\underline{H}}=\left[\left(\mathbf{H}\mathbf{M}_{0,\,0}\right)^{{\rm T}},\,\ldots,\,\left(\mathbf{H}\mathbf{M}_{0,\, k}\right)^{{\rm T}}\right]^{{\rm T}}$,
and
\[
\underline{\mathbf{R}}=\begin{bmatrix}\mathbf{R}_{0} & 0 & \cdots & 0\\
0 & \mathbf{R}_{1} & \ddots & \vdots\\
\vdots & \ddots & \ddots & 0\\
0 & \cdots & 0 & \mathbf{R}_{k}
\end{bmatrix},
\]
 the 4D-Var problem (\ref{eq:4}) further reduces to minimization
of the following cost function:
\begin{equation}
\mathcal{J}_{4D}(\mathbf{x}_{0})=\frac{1}{2}\left\Vert \underline{\mathbf{y}}-\underline{\mathbf{H}}\mathbf{x}_{0}\right\Vert _{\mathbf{\underline{\mathbf{R}}}^{-1}}^{2}+\frac{1}{2}\left\Vert \mathbf{x}_{0}^{b}-\mathbf{x}_{0}\right\Vert _{\mathbf{B}^{-1}}^{2}.\label{eq:5}
\end{equation}

Clearly, (\ref{eq:5}) is a smooth quadratic function of the initial
state of interest $\mathbf{x}_{0}$. Therefore, by setting the derivative
to zero, it has the following analytic minimizer as the analysis state,
\begin{equation}
\mathbf{x}_{0}^{a}=\left(\underline{\mathbf{H}}^{{\rm T}}\underline{\mathbf{R}}^{-1}\underline{\mathbf{H}}+\mathbf{B}^{-1}\right)^{-1}\left(\underline{\mathbf{H}}^{{\rm T}}\underline{\mathbf{R}}^{-1}\underline{\mathbf{y}}+\mathbf{B}^{-1}\mathbf{x}_{0}^{b}\right).\label{eq:6}
\end{equation}

Throughout this study, we used Matlab built-in function \texttt{pcg.m},
described by \citet{[BaiETAL87]}, for obtaining classic solutions
of the 4D-Var in equation (\ref{eq:6}).

Accordingly, it is easy to see \citep[ e.g.,][p.39]{[Dal93]} that
the analysis error covariance is the inverse of the Hessian of (\ref{eq:5}),
as follows:
\begin{equation}
\mathbb{E}\left[\left(\mathbf{x}_{0}^{a}-\mathbf{x}_{0}\right)\left(\mathbf{x}_{0}^{a}-\mathbf{x}_{0}\right)^{{\rm T}}\right]=\left(\underline{\mathbf{H}}^{{\rm T}}\underline{\mathbf{R}}^{-1}\underline{\mathbf{H}}+\mathbf{B}^{-1}\right)^{-1}.\label{eq:7}
\end{equation}

It can be shown that the analysis in the above classic 4D-Var is the
conditional expectation of the true state given observations and the
background state. In other words, the analysis in the classic 4D-Var
problem is the unbiased minimum mean squared error (MMSE) estimator
of the true state \citep[chap.4]{[Lev08]}.

\section{Regularized Variational Data Assimilation}

\subsection{Background}

As is evident, when the Hessian (i.e., $\underline{\mathbf{H}}{}^{{\rm T}}\underline{\mathbf{R}}^{-1}\underline{\mathbf{H}}+\mathbf{B}^{-1}$)
in the classic VDA cost function in (\ref{eq:5}) is ill-conditioned,
the VDA solution is likely to be unstable with large estimation uncertainty.
To study the stabilizing role of the background error, motivated by
the well-known relationship between the Tikhonov regularization and
spectral filtering \citep[e.g.,][]{[GolHO99]}, \citet{[JohNH05],[JohHN06]}
proposed to reformulate the classic VDA problem analogous to the standard
form of the Tikhonov regularization \citep{[Tik77]}. Accordingly,
using a change of variable $\mathbf{z}_{0}=\mathbf{C}_{{\rm B}}^{-1/2}\left(\mathbf{x}_{0}-\mathbf{x}_{0}^{b}\right)$,
letting $\mathbf{B}=\sigma_{b}^{2}\mathbf{C}_{{\rm B}}$ and $\underline{\mathbf{R}}=\sigma_{r}^{2}\mathbf{\underline{\mathbf{C}}}_{{\rm R}}$
, where $\mathbf{C}_{{\rm B}}$ and $\underline{\mathbf{C}}_{{\rm R}}$
are the correlation matrices, the classic variational cost function
was proposed to be reformulated as follows:
\begin{equation}
\mathcal{J}_{4D}(\mathbf{z}_{0})=\left\Vert \mathbf{f}-\mathbf{Gz}_{0}\right\Vert _{2}^{2}+\mu\left\Vert \mathbf{z}_{0}\right\Vert _{2}^{2}.\label{eq:8}
\end{equation}
where the $\ell_{2}$-norm is $\left\Vert \mathbf{x}\right\Vert _{2}=\left(\Sigma_{i=1}^{m}x_{i}^{2}\right)^{1/2}$,
$\mu=\sigma_{r}^{2}/\sigma_{b}^{2}$, $\mathbf{G}=\underline{\mathbf{C}}_{{\rm R}}^{-1/2}\mathbf{\underline{H}}\mathbf{C}_{{\rm B}}^{1/2}$,
and $\mathbf{f}=\underline{\mathbf{C}}_{{\rm R}}^{-1/2}\left(\underline{\mathbf{y}}-\mathbf{\underline{\mathbf{H}}}\mathbf{x}_{0}^{b}\right)$.
Hence, by solving
\[
\mathbf{z}_{0}^{a}={\rm \underset{\mathbf{z}_{0}}{argmin}}\left\{ \mathcal{J}_{4D}(\mathbf{z}_{0})\right\} ,
\]
the analysis can be obtained as, $\mathbf{x}_{0}^{a}=\mathbf{x}_{0}^{b}+\mathbf{C}_{{\rm B}}^{1/2}\mathbf{z}_{0}^{a}$.
Having the above reformulated problem, \citep{[JohHN06]} provided
new insights into the role of the background error covariance matrix
on improving condition number and thus stability of the classic VDA
problem.

To tackle data assimilation of sharp fronts, following the above reformulation,
\citet{[FreNB12]} suggested to add the smoothing $\ell_{1}$-norm
regularization as follows:
\begin{equation}
\mathbf{z}_{0}^{a}=\underset{\mathbf{z}_{0}}{{\rm argmin}}\left\{ \mathcal{J}_{R4D}(\mathbf{z}_{0})+\lambda\left\Vert \mathbf{\Phi}\left(\mathbf{C}_{{\rm B}}^{1/2}\mathbf{z}_{0}+\mathbf{x}_{0}^{b}\right)\right\Vert _{1}\right\} ,\label{eq:9}
\end{equation}
where the $\ell_{1}$-norm is $\left\Vert \mathbf{x}\right\Vert _{1}=\Sigma_{i=1}^{m}\left|x_{i}\right|$;
the non-negative $\lambda$ is called the regularization parameter;
and $\mathbf{\Phi}$ is proposed to be an approximate first-order
derivative operator as follows:
\[
\mathbf{\Phi}=\begin{bmatrix}-1 & 1 &  & 0\\
 & \ddots & \ddots\\
0 &  & -1 & 1
\end{bmatrix}\in\mathbb{R}^{(m-1)\times m}.
\]
Notice that problem (\ref{eq:9}) is a non-smooth optimization as
the derivative of the cost function does not exist at the origin.
\citet{[FreNB12]} recast this problem into a quadratic programing
(QP) with both equality and inequality constraints where the dimension
of the proposed QP is three times larger than that of the original
problem. It is also worth noting that, the reformulations in (\ref{eq:8})
and (\ref{eq:9}) assume that the error covariance matrices are stationary
(i.e., $\mathbf{B}=\sigma_{b}^{2}\mathbf{C}_{{\rm B}}$,\textbf{ $\mathbf{R}=\sigma_{r}^{2}\mathbf{C}_{{\rm R}}$})
and the error variance is distributed uniformly across all of the
problem dimensions. However, without loss of generality, a covariance
matrix $\mathbf{B}\in\mathbb{R}^{m\times m}$ can be decomposed as
$\mathbf{B}={\rm diag\left(\mathbf{s}\right)\,\mathbf{C}_{B}}\,{\rm diag}\left(\mathbf{s}\right)$,
where $\mathbf{s}\in\mathbb{R}^{m}$ is the vector of standard deviations
\citep{[BarMM00]}. Therefore, while one can have an advantage in
stability of computation in (\ref{eq:8}) and (\ref{eq:9}), the stationarity
assumptions and computations of the square roots of the error correlation
matrices might be restrictive in practice.

In the subsequent sections, beyond $\ell_{1}$ regularization of the
first order derivative coefficients, we present a generalized framework
to regularize the VDA problem in a properly chosen transform domain
or basis (e.g., wavelet, Fourier, DCT). The presented formulation
includes smoothing $\ell_{1}$ and $\ell_{2}$-norm regularization
as two especial cases and does not require any explicit assumption
about the stationarity of the error covariance matrices. We recast
the $\ell_{1}$-norm regularized variational data assimilation (RVDA)
into a QP with lower dimension and simpler constraints compared to
the presented formulation by \citet{[FreNB12]}. Furthermore, we introduce
an efficient gradient-based optimization method, suitable for large
scale data assimilation problems. Some results are presented via assimilating
low-resolution and noisy observations into the linear advection-diffusion
equation in a 4D-Var setting.

\subsection{A Generalized Framework to Regularize Variational Data Assimilation
in Transform Domains}

In a more general setting, to regularize the solution of the classic
VDA problem, one may constrain the magnitude of the analysis in the
norm sense as follows:
\begin{eqnarray}
\mathbf{x}_{0}^{a} & = & \underset{\mathbf{x}_{0}}{{\rm argmin}}\left\{ \mathcal{J}_{R4D}(\mathbf{x}_{0})\right\} \nonumber \\
 & {\rm {\rm s.t.}} & \left\Vert \mathbf{\Phi}\mathbf{x}_{0}\right\Vert _{p}^{p}\leq{\rm const.}
\end{eqnarray}
where $\mathbf{\Phi}\in\mathbb{R}^{m\times m}$ is any appropriately
chosen linear transformation, and the $\ell_{p}$-norm is $\left\Vert \mathbf{x}\right\Vert _{p}=\left(\Sigma\left|x_{i}\right|^{p}\right)^{1/p}$
with $p>0$. By constraining the $\ell_{p}$-norm of the analysis,
we implicitly make the solution more stable. In other words, we bound
the magnitude of the analysis state and reduce the instability of
the solution due to the potential ill-conditioning of the classic
cost function. Using the theory of Lagrange multipliers, the above
constrained problem can be turned into the following unconstrained
one:
\begin{equation}
\mathbf{x}_{0}^{a}=\underset{\mathbf{x}_{0}}{{\rm argmin}}\left\{ \frac{1}{2}\left\Vert \underline{\mathbf{y}}-\underline{\mathbf{H}}\mathbf{x}_{0}\right\Vert _{\underline{\mathbf{R}}^{-1}}^{2}+\frac{1}{2}\left\Vert \mathbf{x}_{0}^{b}-\mathbf{x}_{0}\right\Vert _{\mathbf{B}^{-1}}^{2}+\lambda\left\Vert \mathbf{\Phi}\mathbf{x}_{0}\right\Vert _{p}^{p}\right\} .\label{eq:11}
\end{equation}
where the non-negative $\lambda$ is the Lagrange multiplier or regularization
parameter. As is evident, when $\lambda$ tends to zero the regularized
analysis tends to the classic analysis in (\ref{eq:6}), while larger
values are expected to produce more stable solutions but with less
fidelity to the observations and background state. Therefore, in problem
(\ref{eq:11}), the regularization parameter $\lambda$ plays an important
trade-off role and ensures that the magnitude of the analysis is constrained
in the norm sense while keeping it sufficiently close to observations
and background state. Notice that although in special cases there
are some heuristic approaches to find an optimal regularization parameter
\citep[e.g.,][]{[Han93],[JohNH05]}, typically this parameter is selected
empirically based on the problem at hand.

It is important to note that, from the probabilistic point of view,
the regularized problem (\ref{eq:11}) can be viewed as the maximum
a posteriori (MAP) Bayesian estimator. Indeed, the constraint of regularization
refers to the \emph{prior} knowledge about the probabilistic distribution
of the state as $p\left(\mathbf{x}\right)\propto\exp\left(-\lambda\left\Vert \mathbf{\Phi x}\right\Vert _{p}^{p}\right)$.
In other words, we implicitly assume that under the chosen transformation
$\mathbf{\Phi}$ the state of interest can be well explained by the
family of multivariate Generalized Gaussian Density \citep[e.g.,][]{[Nad05]}
which includes the multivariate Gaussian ($p=2$) and Laplace ($p=1$)
densities as special cases. As is evident, because the prior term
is not Gaussian, the posterior density of the above estimator does
not remain in the Gaussian domain and thus characterization of the
a posteriori covariance is not straightforward in this case.

From an optimization view point, the above RVDA problem is convex
with a unique global solution (analysis) when $p\geq1$; otherwise,
it may suffer from multiple local minima. For the special case of
the Gaussian prior ($p=2$) the problem is smooth and resembles the
well-known smoothing norm Tikhonov regularization \citep{[Tik77],[Han10]}.
However, for the case of the Laplace prior ($p=1$) the problem is
non-smooth, and it has received a great deal of attention in recent
years for solving sparse ill-posed inverse problems \citep[see,][and references there in]{[Ela10]}.
It turns out that the $\ell_{1}$-norm regularization promotes sparsity
in the solution. In other words, using this regularization, it is
expected that the number of non-zero elements of $\mathbf{\Phi}\mathbf{x}_{0}^{a}$
be significantly less than the observational dimension. Therefore,
if we know a priori that a specific $\mathbf{\Phi}$ projects a large
number of elements of the state variable of interest onto (near) zero
values, the $\ell_{1}$-norm is a proper choice of the regularization
term that can yield improved estimates of the analysis state \citep[e.g.,][]{[CheDS01],[CanT06],[Ela10]}.

In the subsequent sections, we focus on the 4D-Var problem under the
$\ell_{1}$-norm regularization as follows:

\begin{equation}
\mathbf{x}_{0}^{a}=\underset{\mathbf{x}_{0}}{{\rm argmin}}\left\{ \frac{1}{2}\left\Vert \underline{\mathbf{y}}-\underline{\mathbf{H}}\mathbf{x}_{0}\right\Vert _{\mathbf{\bar{R}}^{-1}}^{2}+\frac{1}{2}\left\Vert \mathbf{x}_{0}^{b}-\mathbf{x}_{0}\right\Vert _{\mathbf{B}^{-1}}^{2}+\lambda\left\Vert \mathbf{\Phi}\mathbf{x}_{0}\right\Vert _{1}\right\} .\label{eq:12}
\end{equation}

It is important to note that the presented formulation in (\ref{eq:12})
shares the same solution with the problem in (\ref{eq:9}) while in
a more general setting, it can handle non-stationary error covariance
matrices and does not require additional computational cost to obtain
their square roots.

\subsubsection{Solution Method via Quadratic Programing}

Due to the separability of the $\ell_{1}$-norm, one of the most well-known
methods, often called basis pursuit \citep[see,][]{[CheDS98],[FigNW07]},
can be used to recast the $\ell_{1}$-norm RVDA problem in (\ref{eq:12})
to a constrained quadratic programming. Here, let us assume that $\mathbf{c}_{0}=\mathbf{\Phi}\mathbf{x}_{0}$,
where $\mathbf{x}_{0}$ and $\mathbf{c}_{0}$ are in $\mathbb{R}^{m}$
and split $\mathbf{c}_{0}$ into its positive $\mathbf{u}_{0}=\max\left(\mathbf{c}_{0},\,0\right)$
and negative $\mathbf{v}_{0}=\max\left(-\mathbf{c}_{0},\,0\right)$
components such that $\mathbf{c}_{0}=\mathbf{u}_{0}-\mathbf{v}_{0}$.
Having this notation, we can express the $\ell_{1}$-norm via a linear
inner product operation as $\left\Vert \mathbf{c}_{0}\right\Vert _{1}=\mathbf{1}_{2m}^{{\rm T}}\mathbf{w}_{0}$,
where $\mathbf{1}_{2m}=[1,\,\ldots,1]^{{\rm T}}\in\mathbb{R}^{2m}$
and $\mathbf{w}_{0}=[\mathbf{u}_{0}^{{\rm T}},\,\mathbf{v}_{0}^{{\rm T}}]^{{\rm T}}$.
Thus, problem (\ref{eq:12}) can be recast as a smooth constrained
quadratic programing problem on non-negative orthant as follows:
\begin{align}
\underset{{\rm \mathbf{w}_{0}}}{{\rm minimize}}\,\,\, & \left\{ \frac{1}{2}\mathbf{w}_{0}^{{\rm T}}\begin{bmatrix}\begin{array}{rr}
\mathbf{Q} & -\mathbf{Q}\\
-\mathbf{Q} & \mathbf{Q}
\end{array}\end{bmatrix}\mathbf{w}_{0}+\left(\lambda\mathbf{1}_{2m}+\begin{bmatrix}\begin{array}{r}
\mathbf{b}\\
-\mathbf{b}
\end{array}\end{bmatrix}\right)^{{\rm T}}\mathbf{w}_{0}\right\} \nonumber \\
 & \,\,{\rm s.t.}\,\,\,\,\,\,\mathbf{w}_{0}\succcurlyeq0,\label{eq:13}
\end{align}
where, $\mathbf{Q}=\mathbf{\Phi}^{-{\rm T}}\left(\underline{\mathbf{H}}^{{\rm T}}\underline{\mathbf{R}}^{-1}\underline{\mathbf{H}}+\mathbf{B}^{-1}\right)\mathbf{\Phi}^{-1}$,
$\mathbf{b}=-\mathbf{\Phi}^{-{\rm T}}\left(\underline{\mathbf{H}}^{{\rm T}}\underline{\mathbf{R}}^{-1}\mathbf{\underline{\mathbf{y}}}+\mathbf{B}^{-1}\mathbf{x}_{0}^{b}\right)$,
and $\mathbf{w}_{0}\succcurlyeq0$ denotes element-wise inequality.

Clearly, given the solution $\hat{\mathbf{w}}_{0}$ of (\ref{eq:13}),
one can easily retrieve $\hat{\mathbf{c}}_{0}$ and thus the analysis
state is $\mathbf{x}{}_{0}^{a}=\mathbf{\Phi}\hat{\mathbf{c}}_{0}$.

The constraint of the QP problem (\ref{eq:13}) is simpler than the
formulation suggested by \citep{[FreNB12]} and allows us to use efficient
and convergent gradient projection methods \citep[e.g.,][]{[Bert76],[sefGL05],[FigNW07]},
suitable for large-scale VDA problems. The dimension of the above
problem seems twice that of the original problem; however, because
of the existing symmetry in this formulation, the computational burden
remains at the same order as the original classic problem (see, appendix
A). Another important observation is that, choosing an orthogonal
transformation (e.g., orthogonal wavelet, DCT, Fourier) for $\mathbf{\Phi}$
is very advantageous computationally, as in this case $\mathbf{\Phi}^{-1}=\mathbf{\Phi}^{{\rm T}}$.

Conceptually, adding relevant regularization terms, we enforce the
analysis to follow a certain regularity and become more stable \citep{[Han10]}.
Here, by regularity, we refer to a certain degree of smoothness in
the analysis state. For instance if we think of $\mathbf{\Phi}$ as
a first order derivative operator, using the smoothing $\ell_{2}$-norm
regularization ($\lambda\left\Vert \mathbf{\Phi}\mathbf{x}_{0}\right\Vert _{2}^{2}$),
we enforce the energy of the solution's increments to be minimal,
which naturally imposes more smoothness. Therefore, using the smoothing
$\ell_{2}$-norm regularization in a derivative space, is naturally
suitable for continuous and smooth physical states. On the other hand,
for piece-wise smooth physical states with isolated singularities
and jumps, it turns out that the use of the smoothing $\ell_{1}$-norm
regularization ($\lambda\left\Vert \mathbf{\Phi}\mathbf{x}_{0}\right\Vert _{1}$)
in a derivative domain is very advantageous. Using this norm in derivative
space, we implicitly constrain the total variation of the solution
which prevents imposing extra smoothness on the solution. Proper selection
of the smoothing norm and $\mathbf{\Phi}$ may fall into the category
of statistical model selection which is briefly explained in the following
subsections.

As briefly explained previously, more stability of the solution comes
from the fact that we constrain the magnitude of the solution by adding
the regularization term and preventing the solution to blow up due
to the ill-conditioning of the VDA problem \citep[see, e.g.,][]{[Han1998],[JohHN06]}.
In ill-conditioned classic VDA problems, it is easy to see that the
inverse of the Hessian in (\ref{eq:7}) may contain very large elements
which spoil the analysis. However, by regularization and making the
problem well-posed, we shrink the size of the elements of the covariance
matrix and reduce the estimation error. We need to emphasize that
this improvement in the analysis error covariance, naturally comes
at the cost of introducing a small bias in the regularized solution
whose magnitude can be kept small by proper selection of the regularization
parameter $\lambda$ \citep[see, e.g.,][]{[Neu98]}.

It is important to note that, for the smoothing $\ell_{1}$-norm regularization
in (\ref{eq:13}), it is easy to show that the regularization parameter
is bounded as $0<\lambda<\left\Vert \mathbf{b}\right\Vert _{\infty},$
where the infinity-norm is $\left\Vert \mathbf{x}\right\Vert _{\infty}=\max\left(\left|x_{1}\right|,\ldots,\,\left|x_{m}\right|\right)$.
For those values of $\lambda$ greater than the upper bound, clearly
the analysis state in (\ref{eq:13}) is the zero vector with maximum
sparsity (see, appendix A).

\section{Examples on Linear Advection-Diffusion Equation}

\subsection{Problem Statement}

The advection-diffusion equation is a parabolic partial differential
equation with a drift and has fundamental applications in various
areas of applied sciences and engineering. This equation is indeed
a simplified version of the general Navier-Stocks equation for a divergence
free and incompressible Newtonian fluid where the pressure gradient
is negligible. In a general form, this equation for a quantity of
$\mathbf{x}(s,\, t)$ is
\begin{eqnarray}
\frac{\partial\mathbf{x}(s,\, t)}{\partial t}+a(s,\, t)\nabla\mathbf{x}(s,\, t) & = & \epsilon\nabla^{2}\mathbf{x}(s,\, t),\nonumber \\
\mathbf{x}(s,\,0) & = & \mathbf{x}_{0}(s),\label{eq:14}
\end{eqnarray}
where $a(s,\, t)$ represents the velocity and $\epsilon\geq0$ denotes
the viscosity constant.

The linear ($a={\rm const}.$) and inviscid form ($\epsilon=0$) of
(\ref{eq:14}) has been the subject of modeling, numerical simulation,
and data assimilation studies of advective atmospheric and oceanic
flows and fluxes. For example, \citet{[Lin98]} argued that the mechanism
of rain-cell regeneration can be well explained by a pure advection
mechanism, \citet{[Joc06]} found that Tropical Instability Waves
(TIWs) need to be modeled by horizontal advection without involving
any temperature mixing length. The nonlinear inviscid form (e.g.,
Burgers' equation) has been used in the shallow water equation and
has been subject of oceanic and tidal data assimilation studies \citep[e.g.,][]{[BenM82],[Eve94b]}.
The linear and viscid form ($\epsilon>0$) has fundamental applications
in modeling of atmospheric and oceanic mixing \citep[e.g.,][chap. 6]{[SmiSM09],[LanV99],[Joc06]},
land-surface moisture and heat transport \citep[e.g.,][]{[AfsM78],[HuI95],[LidZW97],[LiaWL99]},
surface water quality modeling \citep[e.g.,][chap. 8]{[Cha07]}, and
subsurface mass and heat transfer studies \citep[e.g.,][]{[Fet94]}.

Here, we restrict our consideration only to the linear form and present
a series of test problems to demonstrate the effectiveness of the
$\ell_{1}$-norm RVDA in a 4D-Var setting. It is well understood that
the general solution of the linear viscid form of (\ref{eq:14}) relies
on the principle of superposition of linear advection and diffusion.
In other words, the solution at time $t$ is obtained via shifting
the initial condition by $at$, followed by a convolution with the
fundamental Gaussian kernel as follows:
\begin{equation}
\mathcal{D}(s,\, t)=(4\pi\epsilon t)^{-1/2}\exp\left(\frac{-\left|s\right|^{2}}{4\epsilon t}\right),\label{eq:15}
\end{equation}

where the standard deviation is $\sqrt{2\epsilon t}$. As is evident,
the linear shift of size $at$ also amounts to obtaining the convolution
of the initial condition with a Kronecker delta function as follows:

\begin{equation}
\mathcal{A}\left(s-at\right)=\begin{cases}
1 & \,\, s=at\\
0 & {\rm \,\, otherwise}
\end{cases}.
\end{equation}

\subsection{Assimilation Set Up and Results}

\subsubsection{Prognostic Equation and Observation Model}

It is well understood that (circular) convolution in discrete space
can be constructed as a (circulant) Toeplitz matrix-vector product
\citep[e.g.,][]{[ChaHX07]}. Therefore, in the context of a discrete
advection-diffusion model, the temporal diffusivity and spatial linear
shift of the initial condition can be expressed in a matrix form by
$\mathbf{D}_{0,i}$ and $\mathbf{A}_{0,i}$, respectively. In effect,
$\mathbf{D}_{0,i}$ represents a Toeplitz matrix, for which its rows
are filled with discrete samples of the Gaussian Kernel in (\ref{eq:15}),
while the rows of $\mathbf{A}_{0,i}$ contain a properly positioned
Kronecker delta function.

Thus, for our case, the underlying prognostic equation; i.e., $\mathbf{x}_{i}=\mathbf{M}_{0,i}\,\mathbf{x}_{0}$,
may be expressed as follows:
\begin{equation}
\mathbf{x}_{i}=\mathbf{A}_{0,i}\mathbf{D}_{0,i}\,\mathbf{x}_{0}.\label{eq:17}
\end{equation}
In this study, the low-resolution constraints of the sensing system
are modeled using a linear smoothing filter followed by a down-sampling
operation. Specifically, we consider the following time-invariant
linear measurement operator

\begin{equation}
\mathbf{H}=\frac{1}{4}\begin{bmatrix}1\,1\,1\,1 & 0\,0\,0\,0 & \cdots & 0\,0\,0\,0\\
0\,0\,0\,0 & 1\,1\,1\,1 & \cdots & 0\,0\,0\,0\\
\vdots & \vdots & \vdots & \vdots\\
0\,0\,0\,0 & 0\,0\,0\,0 & \cdots & 1\,1\,1\,1
\end{bmatrix}\in\mathbb{R}^{n\times m},\label{eq:18}
\end{equation}

which maps the higher-dimensional state to a lower-dimensional observation
space. In effect, each observation point is then an average and noisy
representation of the four adjacent points of the true state.

\subsubsection{Initial States}

\begin{figure}[t]
\noindent \begin{centering}
\includegraphics[scale=0.5]{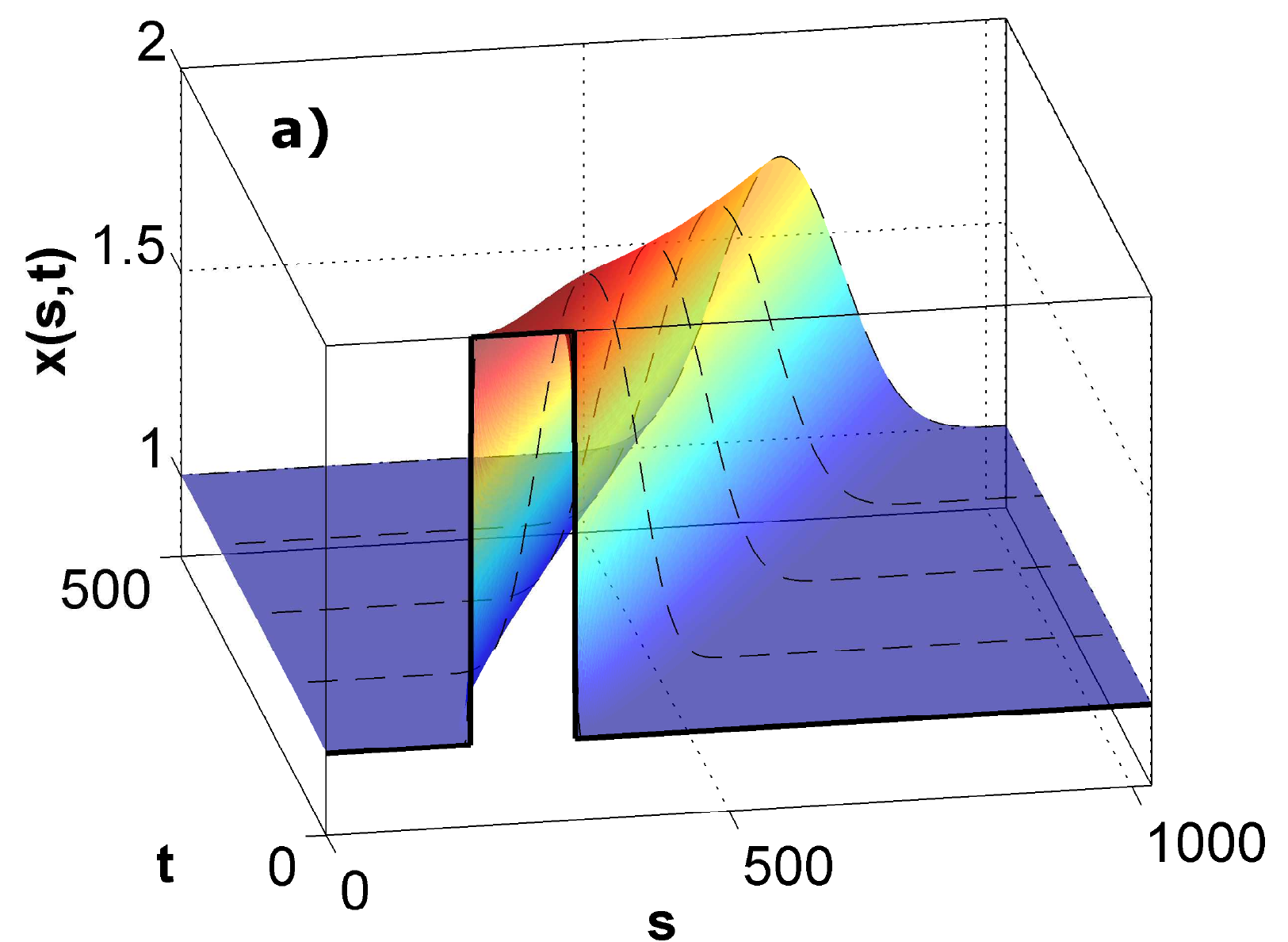}\includegraphics[scale=0.5]{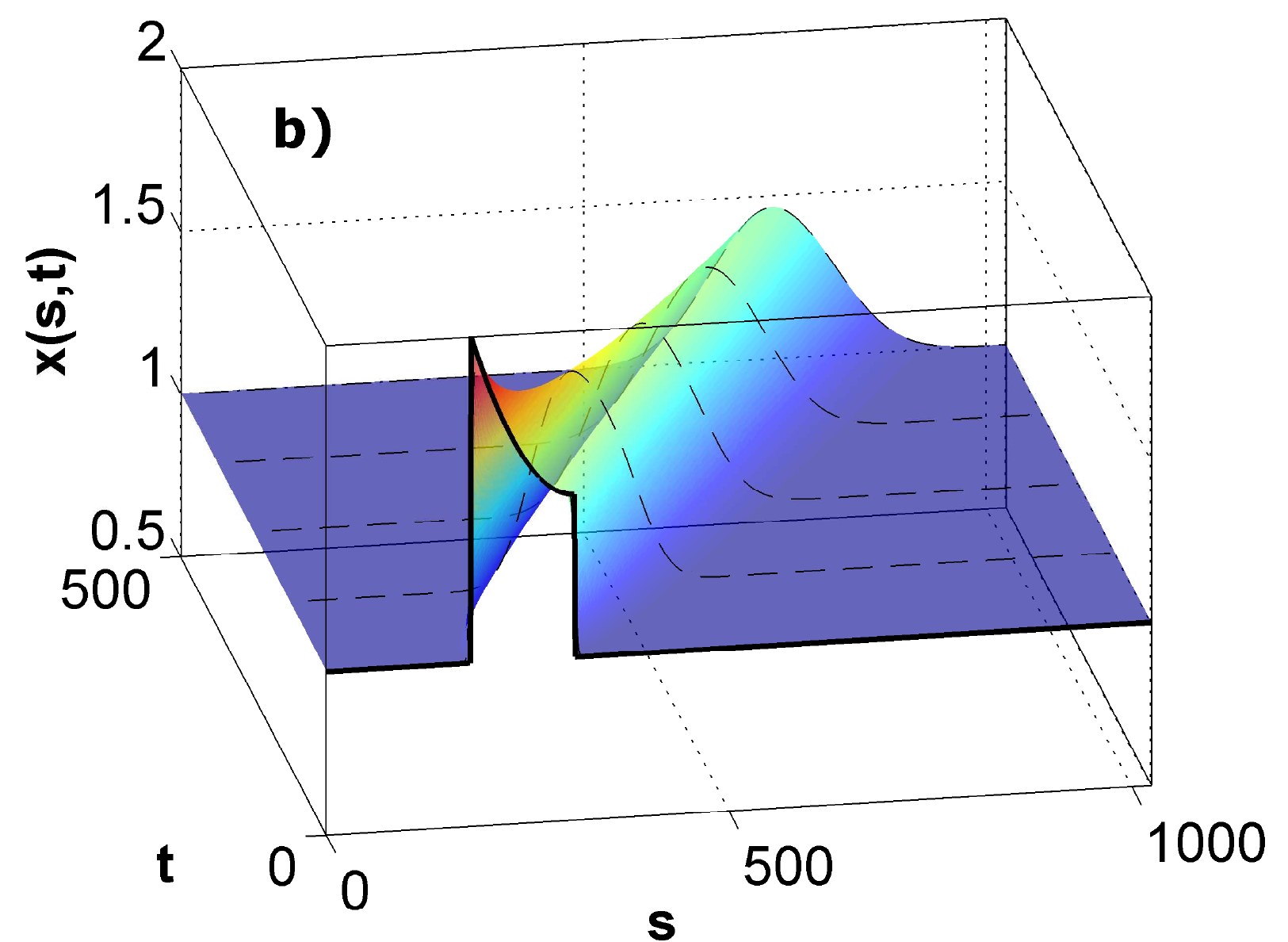}
\par\end{centering}

\noindent \begin{centering}
\includegraphics[scale=0.5]{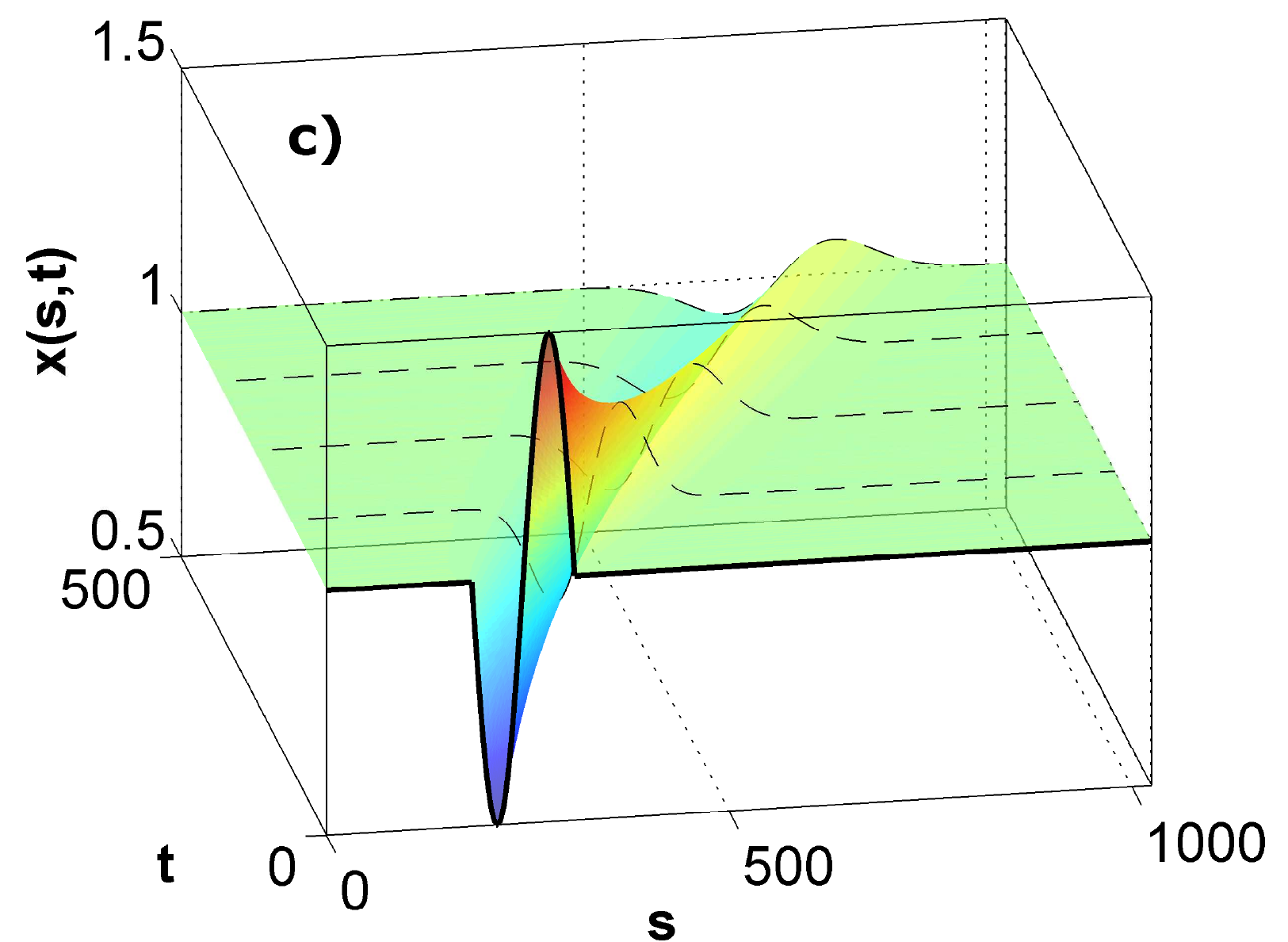}\includegraphics[scale=0.5]{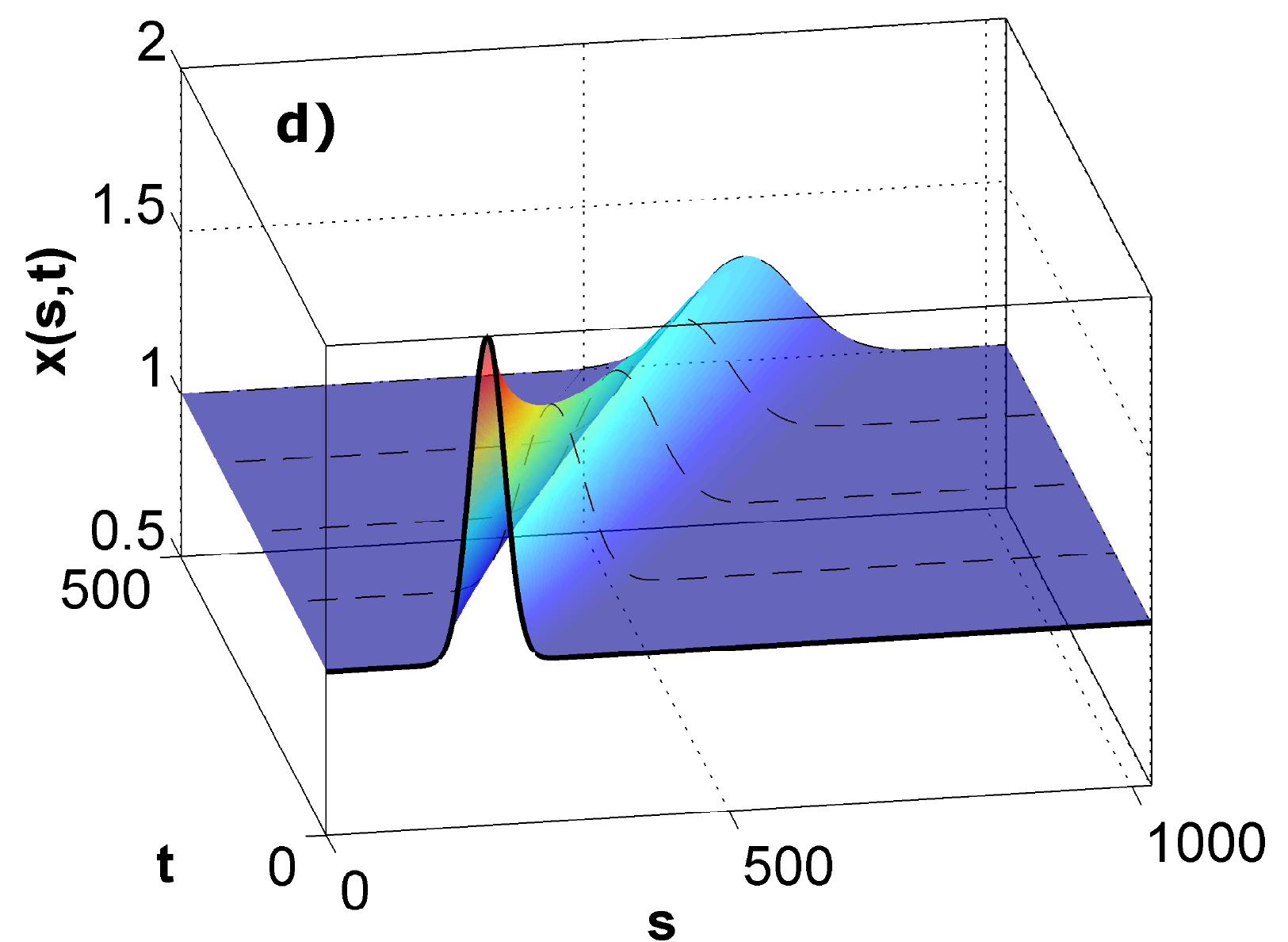}
\par\end{centering}

\caption{Initial conditions and their evolutions with the linear advection-diffusion
equation: (a) flat top-hat (FTH), (b) quadratic top-hat (QTH), (c)
window sinusoid (WS), and (d) squared-exponential (SE). The first
two initial conditions (a, b) exhibit sparse representation in the
wavelet domain while the next two (c, d) show nearly sparse representation
in the discrete cosine domain (DCT). Initial conditions are evolved
under the linear advection-diffusion equation (\ref{eq:14}) with
$\epsilon=4\,[{\rm L}^{2}/{\rm T}]$ and $a=1\,[{\rm L}/{\rm T}]$.
The broken lines show the time instants where the low-resolution and
noisy observations are available in the assimilation interval. \label{Fig:1}}
\end{figure}

\begin{figure}
\noindent \begin{centering}
\includegraphics[scale=0.5]{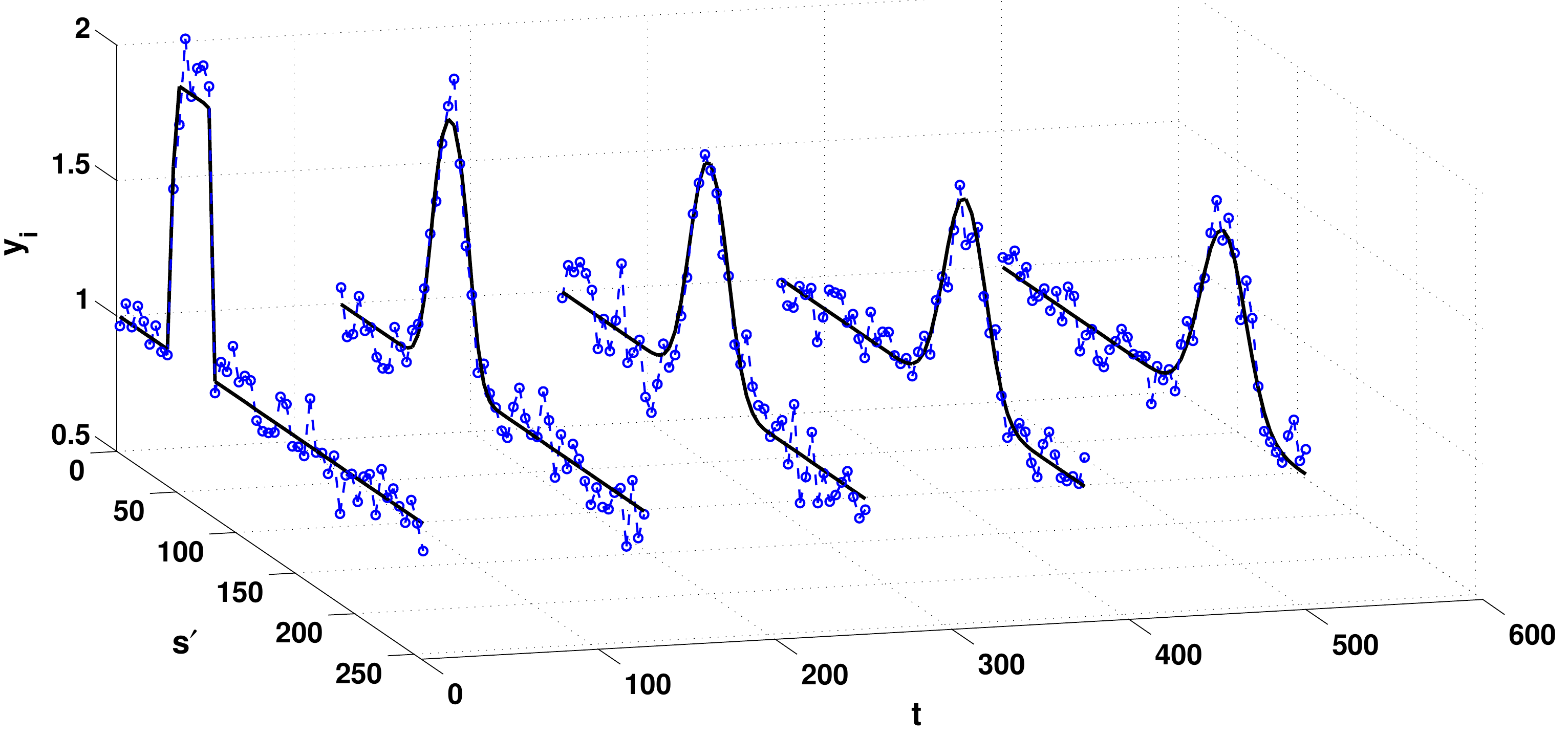}
\par\end{centering}

\caption{A sample representation of the available low-resolution (solid lines)
and noisy observations (broken lines with circles) in every 125 {[}T{]}
time steps in the assimilation window for the flat top-hat initial
condition. Here, the observation error covariance is set to $\mathbf{R}=\sigma_{r}^{2}\mathbf{I}$
with $\sigma_{r}=0.08$ equivalent to ${\rm SNR}=20\,\log\left(\sigma_{\mathbf{x}_{0}}/\sigma_{r}\right)\approx12$
dB. \label{Fig:2} }
\end{figure}

To demonstrate the effectiveness of the proposed $\ell_{1}$-norm
regularization in (\ref{eq:12}), we consider four different initial
conditions which exhibit sparse representation in the wavelet and
DCT domains (Figure \ref{Fig:1}). In particular, we consider: (a)
a flat top-hat, which is a composition of zero-order polynomials and
can be sparsified theoretically using the first order Daubechies wavelet
(DB01) or the Haar basis; (b) a quadratic top-hat which is a composition
of zero and second order polynomials and theoretically can be well
sparsified by wavelets with vanishing moments of order greater than
three \citep[pp.284]{[Mal09]}; (c) a window sinusoid; and (d) a squared
exponential function which exhibits nearly sparse behavior in the
DCT basis. In other words, in the high-frequencies due to the discontinuity
in derivative decay sufficiently fast in the DCT domain. All of the
initial states are assumed to be in $\mathbb{R}^{1024}$ and are evolved
in time with a viscosity coefficient $\epsilon=4\,[{\rm L}^{2}/{\rm T}]$
and velocity $a=1\,[{\rm L}/{\rm T}]$. The assimilation interval
is assumed to be between $0$ and $T=500[{\rm T}]$, where the observations
are sparsely available over this interval at every 125{[}T{]} time
steps (Figure \ref{Fig:1} and \ref{Fig:2}).

\subsubsection{Observation and Background Error}

The observations and background errors are important components of
a data assimilation system that determine the quality and information
content of the analysis. Clearly, the nature and behavior of the errors
are problem-dependent and need to be carefully investigated in a case
by case study. It needs to be stressed that from a probabilistic point
of view, the presented formulation for the $\ell_{1}$-norm RVDA assumes
that both of the error components are unimodal and can be well explained
by the class of Gaussian covariance models. Here, for observation
error, we only consider a stationary white Gaussian measurement error,
$\mathbf{v}\sim\mathcal{N}\left(0,\,\mathbf{R}\right)$, where $\mathbf{R}=\sigma_{r}^{2}\mathbf{I}$
(Figure \ref{Fig:2}).

However, as discussed in \citep{[GasC99]}, the background error can
often exhibit a correlation structure. In this study the first and
second order auto-regressive (AR) Gaussian Markov processes, are considered
for mathematical simulation of a possible spatial correlation in the
background error; see \citet{[GasC99]} for a detailed discussion
about the error covariance models for data assimilation studies.

The AR(1), also known as the Ornestein-Ulenbeck process in infinite
dimension, has an exponential covariance function $\rho(\tau)\propto e^{-\alpha\left|\tau\right|}$.
In this covariance function, $\tau$ denotes the lag either in space
or time, and the parameter $\alpha$ determines the decay rate of
the correlation. The inverse of the correlation decay rate $l_{c}=1/\alpha$
is often called the characteristic correlation length of the process.
The covariance function of the AR(1) model has been studied very well
in the context of stochastic process \citep[e.g.,][]{[Dur99]} and
estimation theory \citep[e.g.,][]{[Lev08]}. For example, it is shown
by \citet[p. 298]{[Lev08]} that the eigenvalues are monotonically
decreasing which may give rise to a very ill-conditioned covariance
matrix in the discrete space, especially for small $\alpha$ or large
correlation length. The covariance function of the AR(2) is more complicated
than the AR(1); however, it has been shown that in special cases,
its covariance function can be explained by $\rho(\tau)\propto e^{-\alpha\left|\tau\right|}\left(1+\alpha\left|\tau\right|\right)$
\citep[p. 31]{[GasC99],[Ste99]}. Note that, both of these covariance
models are stationary and also isotropic as they are only a function
of the magnitude of the correlation lag \citep[pp. 82]{[RasW06]}.
Consequently, the discrete background error covariance is a Hermitian
Toeplitz matrix and can be decomposed into a scalar standard deviation
and a correlation matrix as $\mathbf{B}=\sigma_{b}^{2}\mathbf{C}_{b}$,
where

\[
\mathbf{C}_{b}=\begin{bmatrix}\rho(0) & \rho(1) & \cdots & \rho(m)\\
\rho(1) & \rho(0) & \ddots & \vdots\\
\vdots & \ddots & \ddots & \rho(1)\\
\rho(m) & \cdots & \rho(1) & \rho(0)
\end{bmatrix}\in\mathbb{R}^{m\times m}.
\]

\begin{figure}[t]
\noindent \begin{centering}
\includegraphics[scale=0.68]{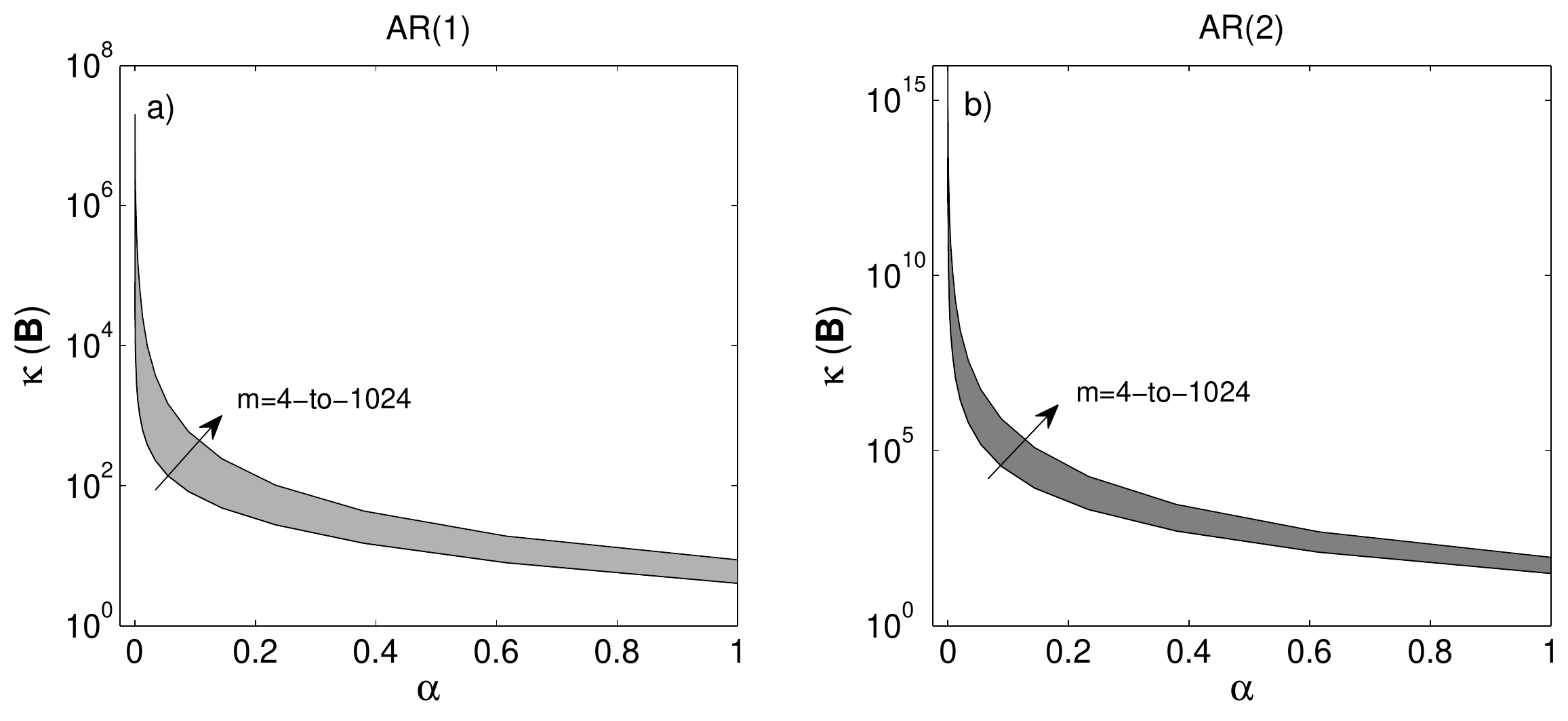}
\par\end{centering}

\caption{Empirical condition numbers of the background error covariance matrices
as a function of parameter $\alpha$ and problem dimension ($m$)
for the AR(1) in (a) and AR(2) in (b). The parameter $\alpha$ varies
along the x-axis and $m$ varies along the different curves of the
condition numbers with values between 4 and 1024. We recall that $\kappa\left(\mathbf{B}\right)$
is the ratio between the largest and smallest singular values of $\mathbf{B}$.
In (a) the covariance matrix is $\mathbf{B}_{ij}=e^{-\alpha\left|i-j\right|}$
and in (b) $\mathbf{B}_{ij}=e^{-\alpha\left|i-j\right|}\left(1+\alpha\left|i-j\right|\right)$,
$1\leq i,\, j\leq m$. It is seen that the condition numbers of the
AR(2) model are significantly larger than those of the AR(1) model
for the same values of the parameter $\alpha$. \label{Fig:3}}
\end{figure}

For the same values of $\alpha$, it is clear that the AR(2) correlation
function decays slower than that of the AR(1). Figure \ref{Fig:3}
shows empirical estimation of the condition number of the reconstructed
correlation matrices at different dimensions ranging from $m=$4 to
1024. As is evident, the error covariance of the AR(2) has a larger
condition number than that of AR(1) for the same value of the parameter
$\alpha$. Clearly, as the background error plays a very important
role on the overall condition number of the Hessian in the cost function
in (\ref{eq:5}), an ill-conditioned background error covariance makes
the solution more unstable with larger uncertainty around the obtained
analysis.

\begin{figure}[t]
\noindent \begin{centering}
\includegraphics[scale=0.65]{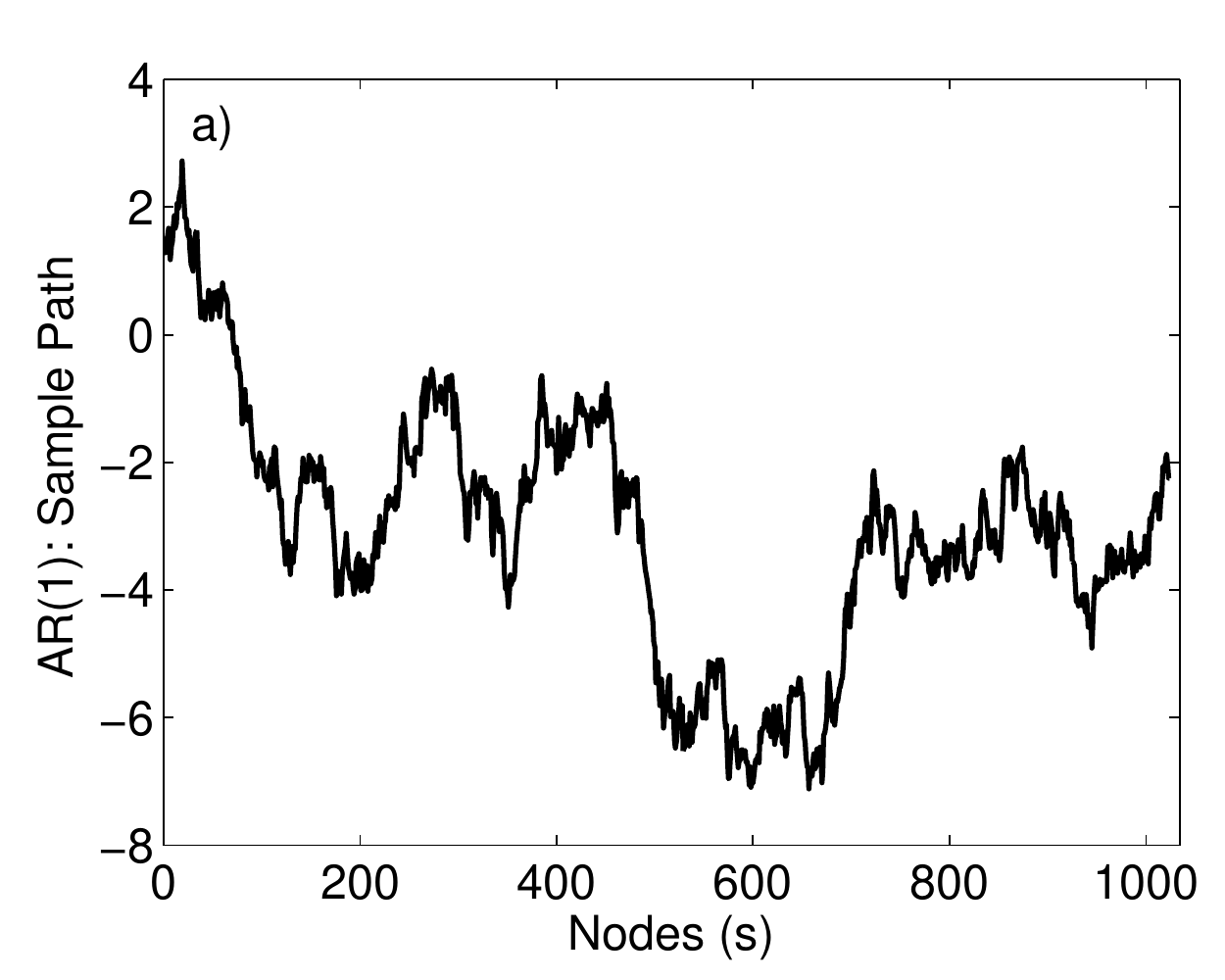}\includegraphics[scale=0.65]{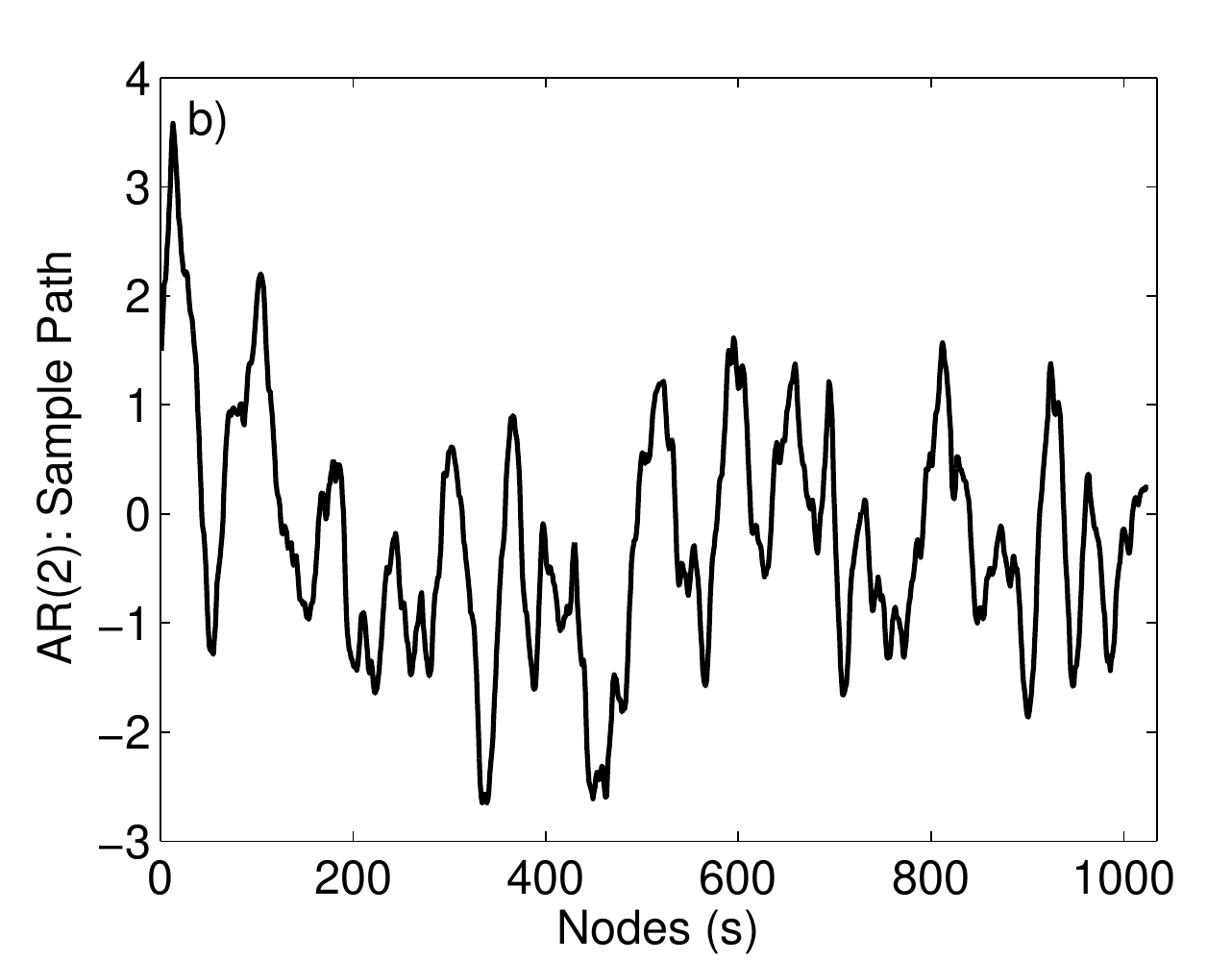}
\par\end{centering}

\caption{Sample paths of the used correlated background error: (a) the sample
path for the AR(1) covariance matrix with $\alpha^{-1}=150$, and
(b) the sample path for the AR(2) covariance matrix with $\alpha^{-1}=25$.
The paths are generated by multiplying a standard white Gaussian noise
$\mathbf{e}\sim\mathcal{N}\left(0,\,\mathbf{I}\right)$ form the left
by the lower triangular matrix $\mathbf{L}$, obtained by Cholesky
factorization of the background error covariance matrix, that is $\mathbf{B}=\mathbf{LL}^{{\rm T}}$.
It is seen that for small $\alpha$, the sample paths exhibit large
scale oscillatory behavior that can potentially corrupt low-frequency
components of the underlying state. \label{Fig:4}}
\end{figure}

Figure \ref{Fig:4} shows a sample path of the chosen error models
for the background error. Generally speaking, a correlated error contains
large-scale (low-frequency) components that can corrupt the main spectral
components of the true state at the same frequency range. Therefore,
this type of error can superimpose with the large-scale characteristic
features of the initial state and its removal is naturally more difficult
than that of the white error via a data assimilation methodology.

\subsection{Results of Assimilation Experiments}

In this subsection, we present the results of the proposed regularized
data assimilation as expressed in equation (\ref{eq:12}). We first
present the results for the white background error and then discuss
the correlated error scenarios. As previously explained, the first
two initial conditions exhibit sharp transitions and are naturally
sparse in the wavelet domain. For those initial states (Figure \ref{Fig:1}a,
b) we have used classic orthogonal wavelet transformation by \citet{[Mal89]}.
Indeed, the columns of $\mathbf{\Phi}\in\mathbb{R}^{1024\times1024}$
in this case contain the chosen wavelet basis that allow us to decompose
the initial state of interest into its wavelet representation coefficients,
as $\mathbf{c}=\mathbf{\Phi}\mathbf{x}$ (forward wavelet transform).
On the other hand, due to the orthogonality of the chosen wavelet
$\mathbf{\Phi}\mathbf{\Phi}^{{\rm T}}=\mathbf{I}$, rows of $\mathbf{\Phi}^{{\rm T}}$
contain the wavelet basis that allows us to reconstruct the initial
state from its wavelet representation coefficients, that is $\mathbf{x}=\mathbf{\Phi}^{{\rm T}}\mathbf{c}$
(inverse wavelet transform). We used a full level of decomposition
without any truncation of wavelet decomposition levels to produce
a fully sparse representation of the initial state. For example, in
our case where $\mathbf{x}\in\mathbb{R}^{1024}$, we have used ten
levels of decomposition.

For the last two initial states (Figure \ref{Fig:1}c, d) we used
DCT transformation \citep[e.g.,][]{[RaoY90]} which expresses the
state of interest by a linear combination of the oscillatory cosine
functions at different frequencies. It is well understood that this
basis has a very strong compaction capacity to capture the energy
content of sufficiently smooth states and sparsely represent them
via a few elementary cosine waveforms. Note that, this transformation
is also orthogonal ($\mathbf{\Phi}\mathbf{\Phi}^{{\rm T}}=\mathbf{I}$)
and contrary to the Fourier transformation, the expansion coefficients
are real.

\subsubsection{White Background Error}

\begin{figure}[t]
\noindent \begin{centering}
\includegraphics[scale=0.45]{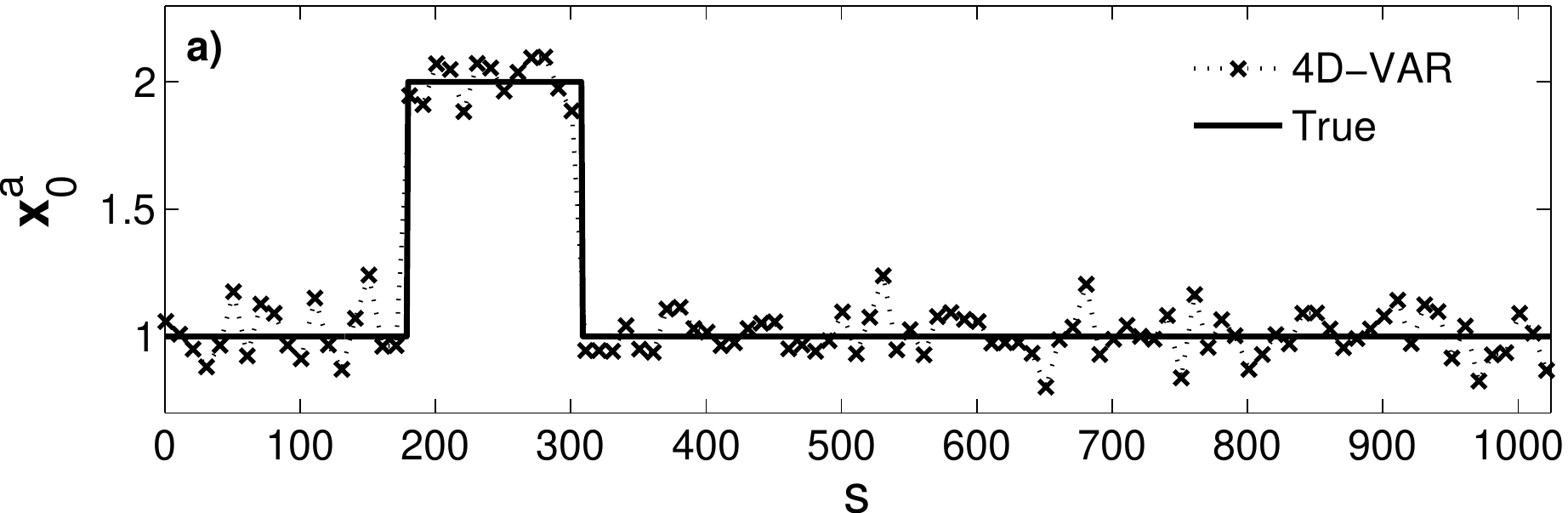}\includegraphics[scale=0.45]{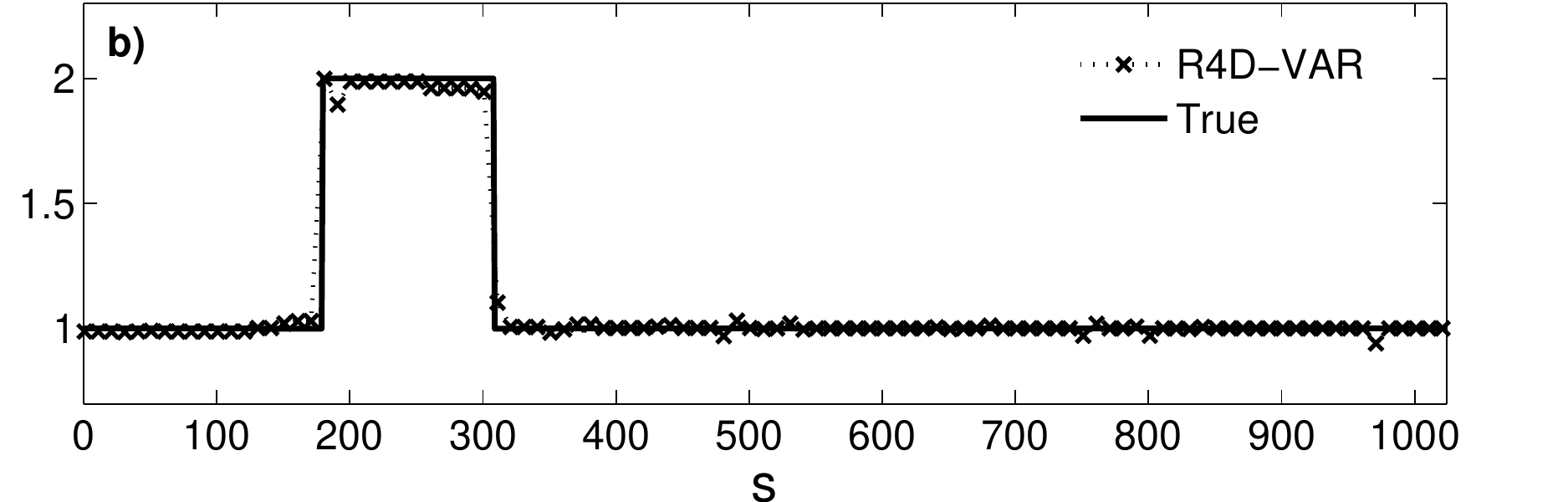}
\par\end{centering}

\noindent \begin{centering}
\includegraphics[scale=0.45]{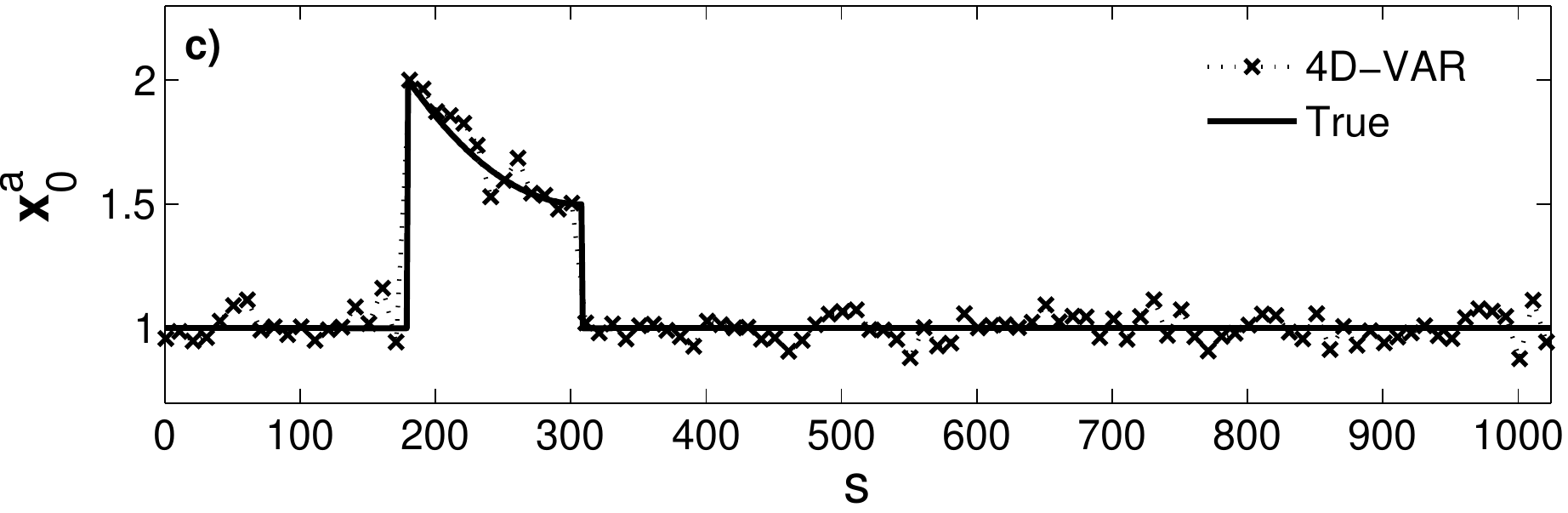}\includegraphics[scale=0.45]{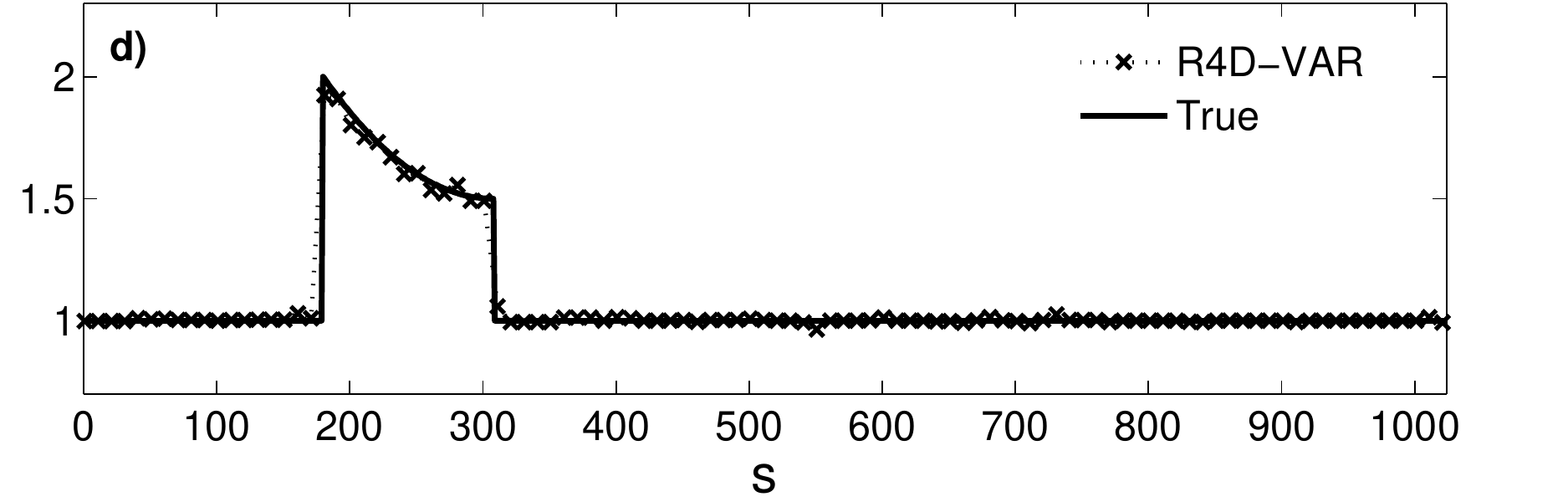}
\par\end{centering}

\noindent \begin{centering}
\includegraphics[scale=0.45]{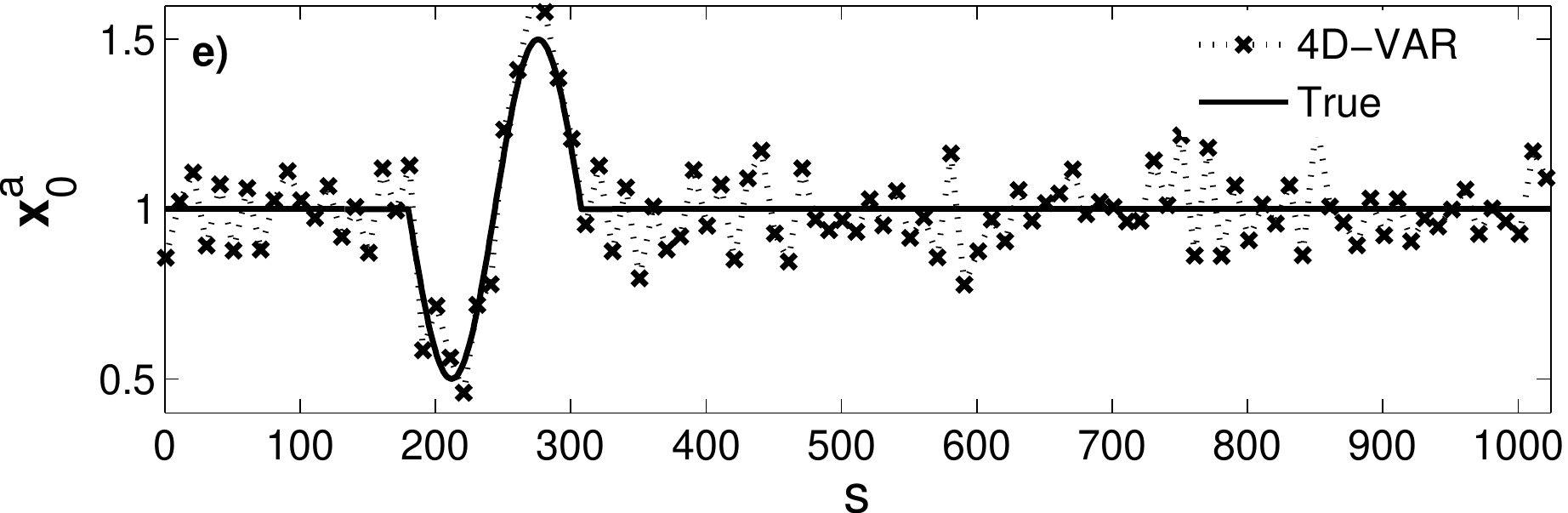}\includegraphics[scale=0.45]{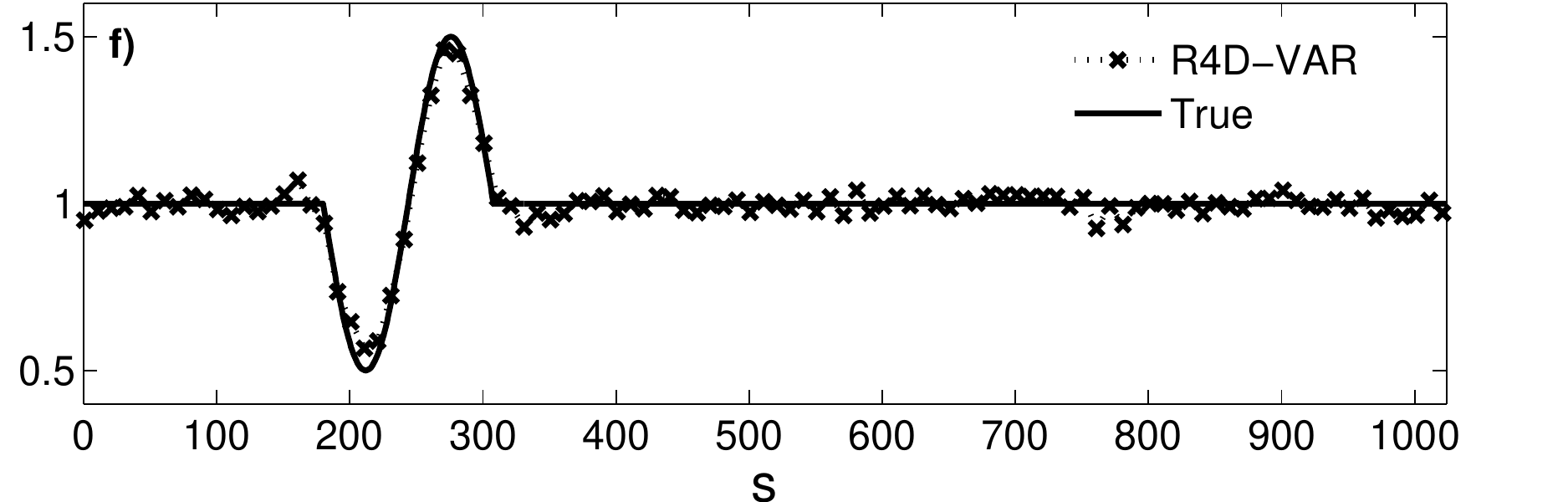}
\par\end{centering}

\noindent \begin{centering}
\includegraphics[scale=0.45]{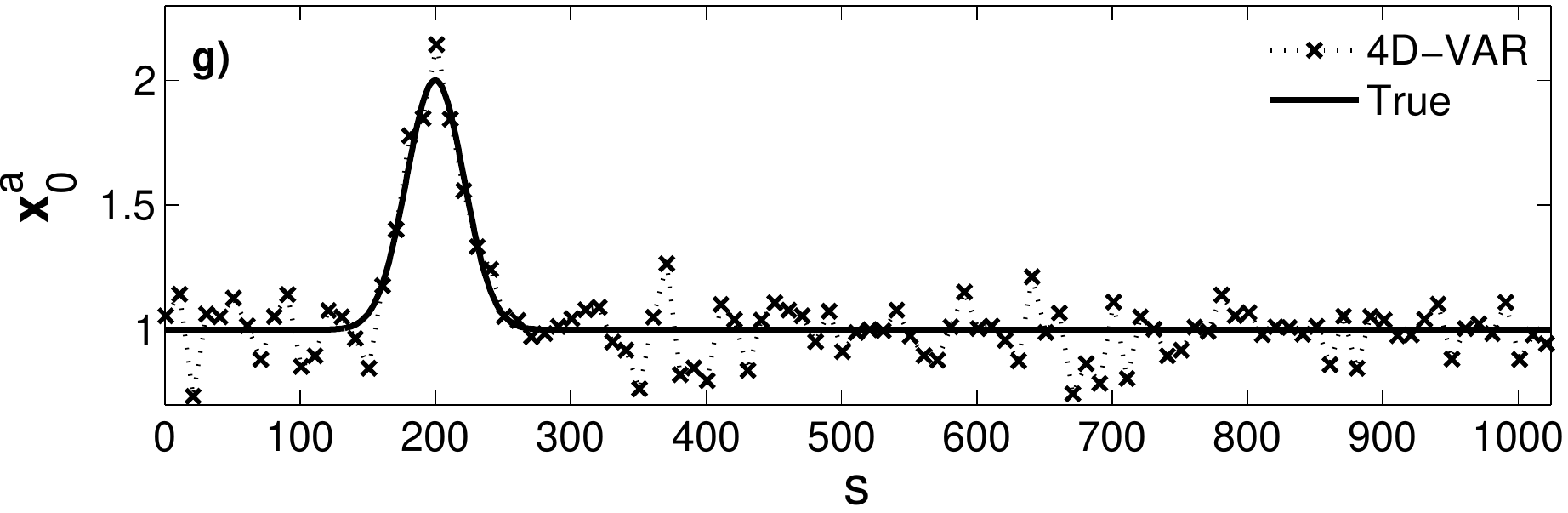}\includegraphics[scale=0.45]{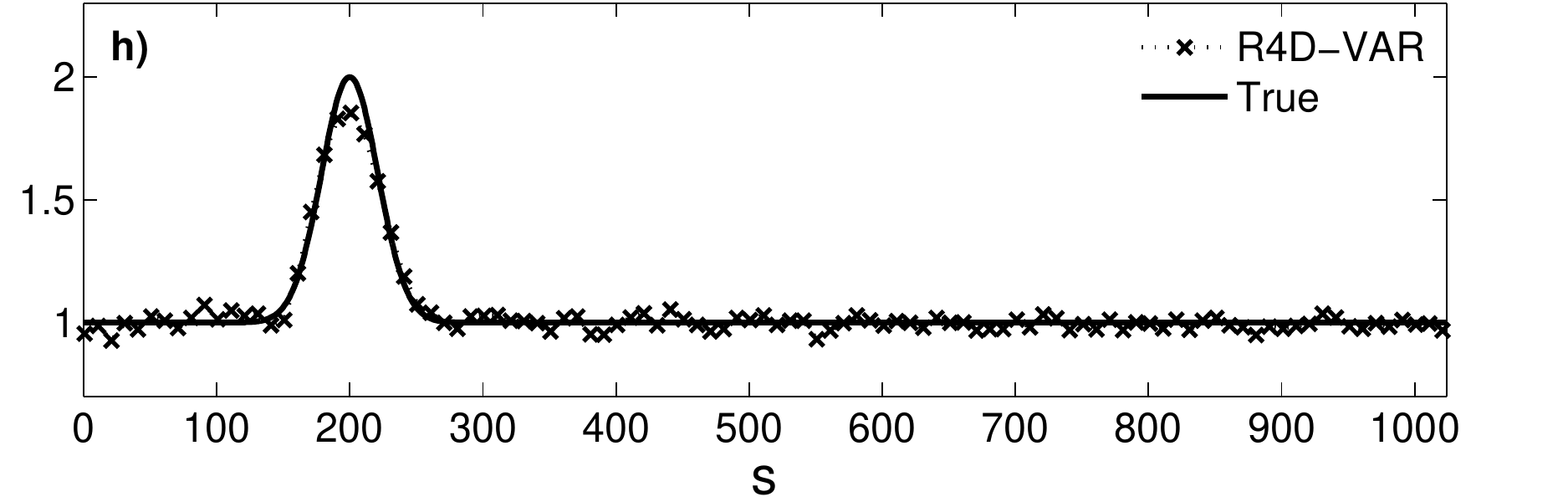}
\par\end{centering}

\caption{The results of the classic 4D-Var (left panel) versus the results
of $\ell_{1}$-norm R4D-Var (right panel) for the tested initial conditions
in a white Gaussian error environment. The solid lines are the true
initial conditions and the crosses represent the recovered initial
states or the analysis. In general, the results of the classic 4D-Var
suffer from overfitting while the background and observation errors
are suppressed and the sharp transitions and peaks are effectively
recovered in the regularized analysis. \label{Fig:5}}
\end{figure}

\noindent For the white background and observation error covariance
matrices ($\mathbf{B}=\sigma_{b}^{2}\mathbf{I}$, $\mathbf{R}=\sigma_{r}^{2}\mathbf{I}$
), we considered $\sigma_{b}=0.10$ (${\rm SNR}\cong10.5$ dB) and
$\sigma_{r}=0.08$ (${\rm SNR}\cong12$ dB), respectively. Some results
are shown in Figure \ref{Fig:5} for the selected initial conditions.
It is clear that the $\ell_{1}$-norm regularized solution markedly
outperforms the classic 4D-Var solutions in terms of the selected
metrics. Indeed, in the regularized analysis the error is sufficiently
suppressed and filtered, while characteristic features of the initial
state are well-preserved. On the other hand, classic solutions typically
over-fitted and followed the background state rather than extracting
the true state. As a result, we can argue that for the white error
covariance the classic 4D-Var has a very weak filtering effect which
is an essential component of an ideal data assimilation scheme. This
over-fitting may be due to the redundant (over-determined) formulation
of the classic 4D-Var; see \citep{[Haw04]} for a general explanation
on overfitting problems in statistical estimators and also see \citet[p.41]{[Dal93]}.

The average of the results for 30 independent runs is reported in
Table \ref{Table1}. Three different lump quality metrics are examined
as follows:
\begin{eqnarray}
{\rm MSE}_{r} & = & \left\Vert \mathbf{x}_{0}^{t}-\mathbf{x}_{0}^{a}\right\Vert _{2}/\left\Vert \mathbf{x}_{0}^{t}\right\Vert _{2}\nonumber \\
{\rm MAE}_{r} & = & \left\Vert \mathbf{x}_{0}^{t}-\mathbf{x}_{0}^{a}\right\Vert _{1}/\left\Vert \mathbf{x}_{0}^{t}\right\Vert _{1}\nonumber \\
{\rm BIAS}_{r} & = & \left|\mathbf{\bar{x}}_{0}^{t}-\mathbf{\bar{x}}_{0}^{a}\right|/\left|\mathbf{\bar{x}}_{0}^{t}\right|\label{eq:19}
\end{eqnarray}

namely, relative mean squared error $\left({\rm MSE}_{r}\right)$,
relative mean absolute error $\left({\rm MAE}_{r}\right)$, and relative
Bias $\left({\rm BIAS}_{r}\right)$. In (\ref{eq:19}) $\mathbf{x}_{0}^{t}$
denotes the true initial condition, $\mathbf{x}_{0}^{a}$ is the analysis,
and upper bar denote the expected value. It is seen that based on
the selected lump quality metrics, the $\ell_{1}$-norm R4D-Var significantly
outperforms the classic 4D-Var. In general, the ${\rm MAE}_{r}$ metric
is improved more than the ${\rm MSE}_{r}$ metric in the presented
experiments. The best improvement is obtained for the flat top-hat
initial condition (FTH), where the sparsity is very strong compared
to the other initial conditions. In other words, the $\ell_{1}$-norm
R4D-Var is more effective for stronger sparsity of the initial state.
The ${\rm MSE}_{r}$ metric is improved almost three orders of magnitude,
while the ${\rm MAE}_{r}$ improvement reaches up to six orders of
magnitude in the FTH initial condition. We need to note that although
the trigonometric functions can be sparsely represented in the DCT
domain, here we used a window sinusoid, which suffers from discontinuities
over the edges and can not be perfectly sparsified in the DCT domain.
However, we see that even in a weaker sparsity, the results of the
$\ell_{1}$-norm R4D-Var are still much better than the classic solution.

\begin{table}
\noindent \begin{centering}
\begin{tabular}{|c|c|c|c|c|c|c|}
\hline
\multicolumn{7}{|c|}{White Background Error}\tabularnewline
\hline
\hline
 & \multicolumn{2}{c|}{${\rm MSE}_{r}$} & \multicolumn{2}{c|}{${\rm MAE}_{r}$} & \multicolumn{2}{c|}{${\rm BIAS}_{r}$}\tabularnewline
\hline
 & R4D-Var & 4D-Var & R4D-Var & 4D-Var & R4D-Var & 4D-Var\tabularnewline
\hline
\textbf{FTH} & 0.0188 & 0.0690 & 0.0099 & 0.0589 & 0.0016 & 0.0004\tabularnewline
\hline
\textbf{QTH} & 0.0152 & 0.0515 & 0.0083 & 0.0414 & 0.0030 & 0.0016\tabularnewline
\hline
\textbf{WS} & 0.0296 & 0.0959 & 0.0229 & 0.0771 & 0.0038 & 0.0022\tabularnewline
\hline
\textbf{SE} & 0.0316 & 0.0899 & 0.0235 & 0.0728 & 0.0018 & 4.26${\rm e}-5$\tabularnewline
\hline
\end{tabular}
\par\end{centering}

\caption{Expected values of the ${\rm MSE}_{r}$, ${\rm MAE}_{r}$, and ${\rm BIAS}_{r}$,
defined in (\ref{eq:19}), for 30 independent runs. The background
and observation errors are white ($\mathbf{B}=\sigma_{b}^{2}\mathbf{I}$,
$\mathbf{R}=\sigma_{r}^{2}\mathbf{I}$ ), where $\sigma_{b}=0.10$
(${\rm SNR}\cong10.5$ dB) and $\sigma_{r}=0.08$ (${\rm SNR}\cong12$
dB). The initial conditions are: flat top-hat (FTH), quadratic top-hat
(QTH), window sinusoid (WS), and squared-exponential (SE). The results
are reported for both the classic 4D-Var and the regularized 4D-Var
(R4D-Var). \label{Table1} }
\end{table}

\subsubsection{Correlated background error}

\begin{figure}[h]
\noindent \begin{centering}
\includegraphics[scale=0.52]{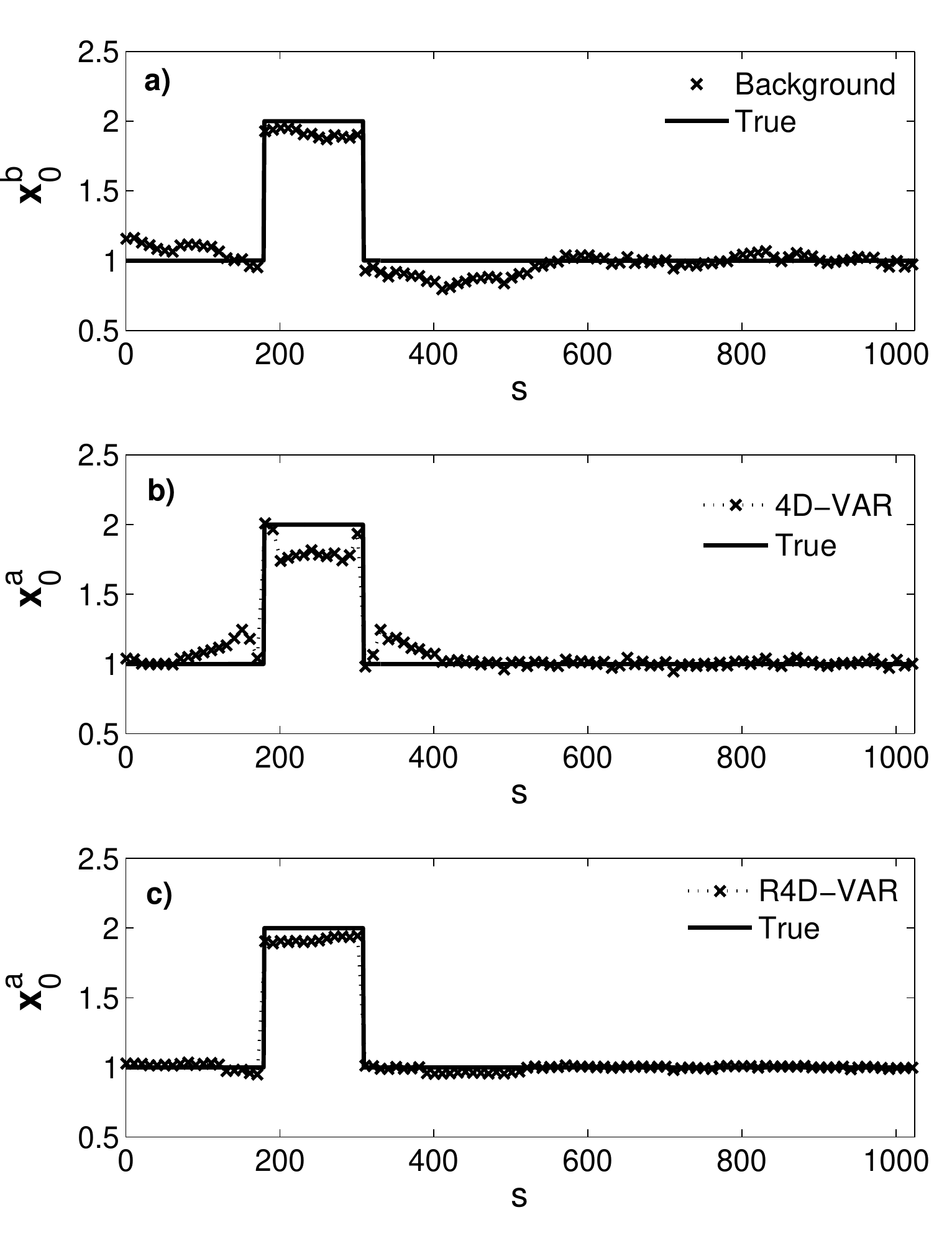}\includegraphics[scale=0.52]{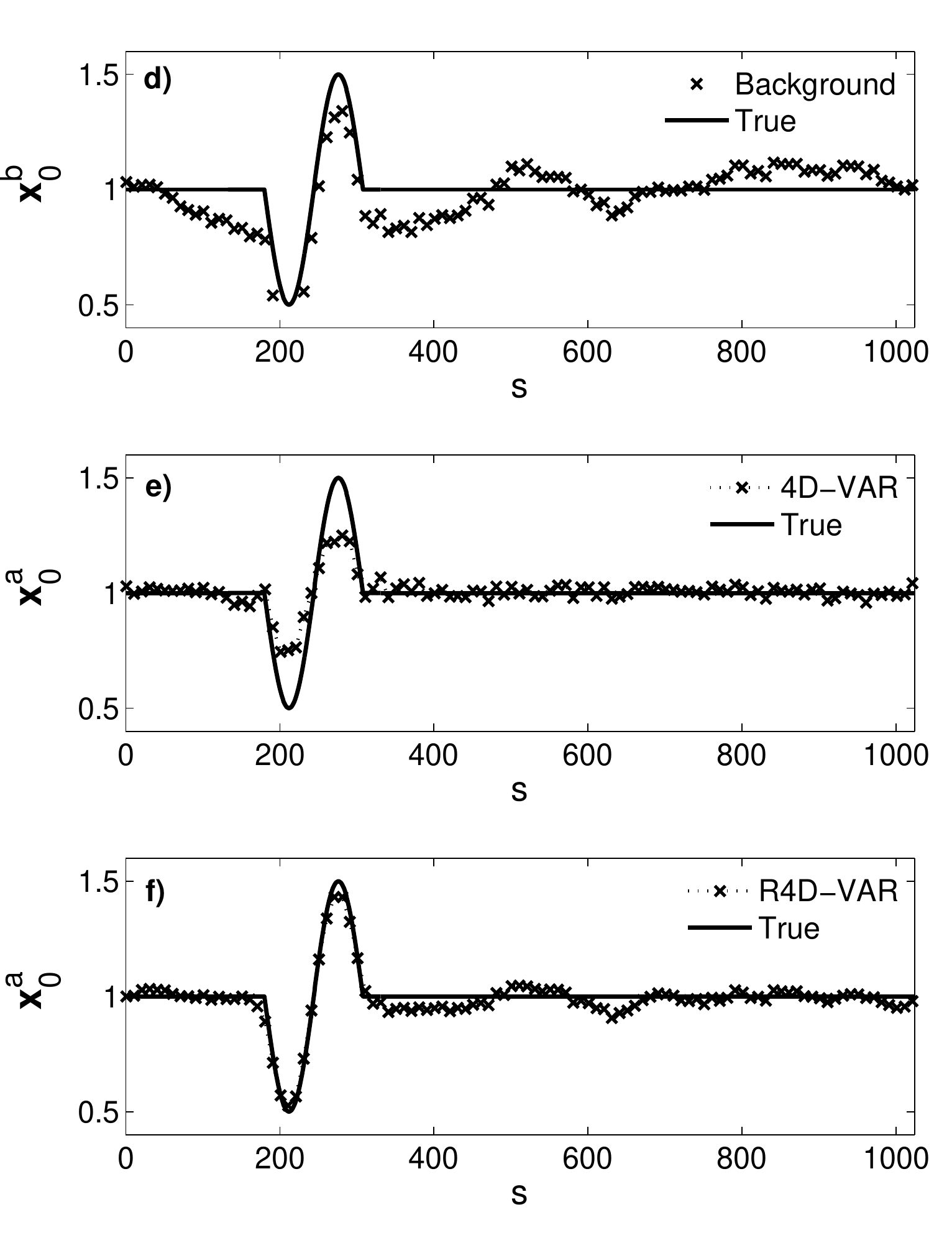}
\par\end{centering}

\caption{Comparison of the results of the classic 4D-Var (b,$\:$e) and $\ell_{1}$-norm
R4D-Var (c,$\:$f) for the top-hat (left panel) and window sinusoid
(right panel) initial conditions. The background states in (a) and
(d) are defined by adding correlated errors using an AR(1) covariance
model of $\rho(\tau)\propto e^{-\alpha\left|\tau\right|}$, where
$\alpha=1/250$. The results show that the $\ell_{1}$-norm R4D-Var
improves recovery of sharp jumps and peaks and results in a more stable
solution compared to the classic 4D-Var; see Figure \ref{Fig:7} for
quantitative results. \label{Fig:6}}
\end{figure}

In this part, the background error $\mathbf{B}=\sigma_{b}^{2}\mathbf{C}_{b}$
is considered to be correlated. As previously discussed, typically
longer correlation length creates ill-conditioning in the background
error covariance matrix and makes the problem more unstable. On the
other hand, the correlated background error covariance imposes smoothness
on the analysis \citep[see,][]{[GasC99]}, improves filtering effects,
and makes the classic solution to be less prone to overfitting. In
this subsection, we examine the effect of correlation length on the
solution of data assimilation and compare the results of the sparsity
promoting R4D-Var with the classic 4D-Var. Here, we do not apply any
preconditioning as the goal is to emphasize on the stabilizing role
of the $\ell_{1}$-norm regularization in the presented formulaiton.
In addition, for brevity, the results are only reported for the top-hat
and window sinusoid initial condition, which are solved in the wavelet
and DCT domains, respectively.
\begin{description}
\item [{a)}] Results for the AR(1) background error
\end{description}
As is evident, in this case, the background state is defined by adding
AR(1) correlated error to the true state (\ref{Fig:6}a,d) which is
known to us for these experimental studies. Figure \ref{Fig:6} demonstrates
that in the case of correlated error the classic 4D-Var is less prone
to overfitting compared to the case of the uncorrelated error in Figure
\ref{Fig:5}. Typically in the flat top-hat initial condition (FTH)
with sharp transitions, the classic solution fails to capture those
sharp jumps and becomes spoiled around those discontnuities (Figure
\ref{Fig:6}b). For the trigonometric initial condition (WS), the
classic solution is typically overly smooth and can not capture the
peaks (Figure \ref{Fig:6}e). These deficiencies in classic solutions
typically become more pronounced for larger correlation lengths and
thus more ill-conditioned problems. On the other hand, the $\ell_{1}$-norm
R4D-Var markedly outperforms the classic method by improving the recovery
of the sharp transitions in FTH and peaks in WS (Figure \ref{Fig:6}).

\begin{figure}[t]
\noindent \begin{centering}
\includegraphics[scale=0.55]{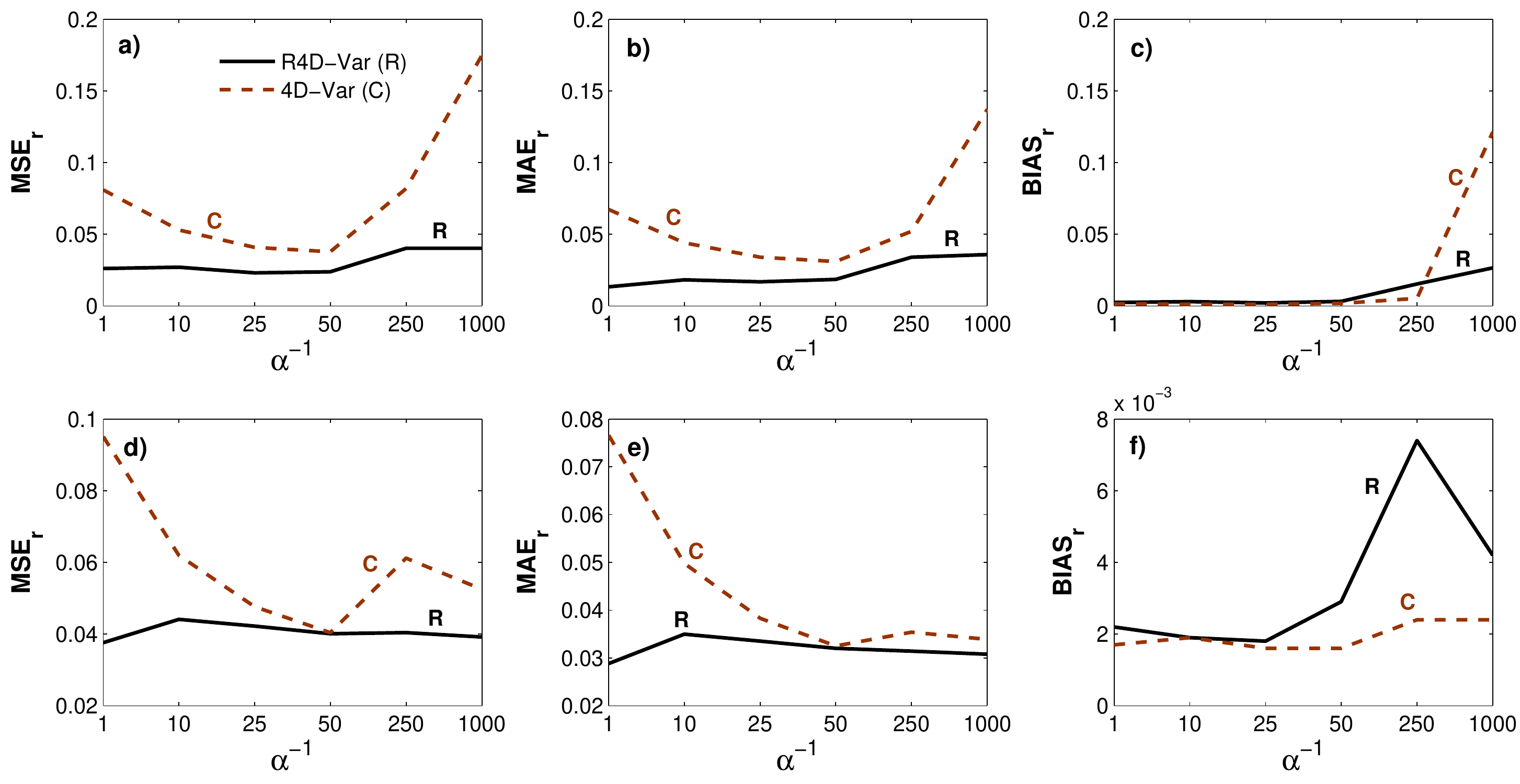}
\par\end{centering}

\caption{Comparison of the results of the proposed $\ell_{1}$-norm R4D-Var
(solid lines) and the classic 4D-Var (broken lines) under the AR(1)
background error for different correlation characteristic length scales
($\alpha^{-1}$). Top panel: (a-c) the chosen quality metrics for
the top-hat initial condition (FTH); Bottom panel: (d-f) the metrics
for the window sinusoid initial condition (WS). These results, averaged
over 30 independent runs, demonstrate significant improvements in
recovering the analysis state by the proposed $\ell_{1}$-norm R4D-Var
compared to the classic 4D-Var. \label{Fig:7}}
\end{figure}

We examined a relatively wide range of applicable correlation lengths,
$\alpha^{-1}\in\left\{ 1,\,10,\,25,\,50,\,250,\,1000\right\} $, which
correspond to decades of variations ranging from $10^{1}$ to $10^{6}$
in the condition number $\kappa\left(\mathbf{B}\right)$ of the background
error covariance matrices (see Figure \ref{Fig:3}a). The assimilation
results using different correlation lengths are demonstrated in Figure
\ref{Fig:7}. To have a robust conclusion about comparison of the
proposed R4D-Var with the classic 4D-Var, the plots in this figure
demonstrate the expected values of the quality metrics for 30 independent
runs.

It can be seen that for small error correlation lengths ($\alpha^{-1}\lesssim25$),
the improvement of the R4D-Var is very significant while in the medium
range ($25\lesssim\alpha^{-1}\lesssim50$) the classic solution becomes
more competitive and closer to the regularized analysis. As previously
mentioned, this improvement in the classic solutions is mainly due
to the smoothing effect of the background covariance matrix. However,
for larger correlation lengths ($\alpha^{-1}\gtrsim50$), the differences
of the two methods are more drastic as the classic solutions become
more unstable and fail to capture the underlying structure of the
initial state of interest. In general, we see that the ${\rm MSE}_{r}$
and ${\rm MAE}_{r}$ metrics are improved for all examined background
error correlation lengths. As expected, the regularized solutions
are slightly biased compared to classic solutions; however, the magnitude
of the bias is not significant compared to the mean value of the initial
state (see Figure \ref{Fig:7}). Figure \ref{Fig:7} also shows a
very important outcome of regularization which implies that the R4D-Var
is almost insensitive to the studied range of correlation length and
thus condition number of the problem. This confirms the stabilizing
role of regularization and needs to be further studied for large scale
and operational data assimilation problems. Another important observation
is that, for extremely correlated background error, the classic R4D-Var
may produce analysis with larger bias than the proposed R4D-Var (Figure
\ref{Fig:7}c). This unexpected result might be due to the presence
of spurious bias in the background state coming from a strongly correlated
error. In other words, a strongly correlated error may shift the mean
value of the background state significantly and create a large bias
in the solution of the classic 4D-Var. In this case, the improved
performance of the R4D-Var may be due to its stronger stability and
filtering properties.

\begin{description}
\item [{b)}] Results for the AR(2) background error
\end{description}
The AR(2) model is suitable for errors with higher order Markovian
structure compared to the AR(1) model. As is seen in Figure (\ref{Fig:4}),
the condition number of the AR(2) covariance matrix is much larger
than the AR(1) for the same values of the parameter $\alpha$ in the
studied covariance models. Here, we limited our experiments to fewer
characteristic correlation lengths of $\alpha^{-1}=\left\{ 1,\,5,\,25,\,50\right\} $.
We constrained our considerations to $\alpha^{-1}\lesssim50$ , because
for larger values (slower correlation decay rates) the condition number
of $\mathbf{B}$ exceeds $10^{8}$ and almost both methods failed
to obtain the analysis without any preconditioning effort.

In our case study, for $\alpha^{-1}\lesssim25$, where $\kappa(\mathbf{B})\lesssim10^{6}$,
the proposed R4D-Var outperforms the 4D-Var similar to what has been
explained for the AR(1) error in the previous subsection. However,
we found that for $25\lesssim\alpha^{-1}\lesssim50$, where $10^{6}\lesssim\kappa(\mathbf{B})\lesssim10^{8}$,
without proper preconditioning, the used conjugate gradient algorithm
fails to obtain the analysis state in the 4D-Var (Table \ref{Table2}).
On the other hand, due to the role of the proposed regularization,
the R4D-Var remains sufficiently stable; however, its effectiveness
deteriorated compared to the cases where the condition numbers were
lower. This observation verifies the known role of the proposed regularization
for improving the condition number of the variational data assimilation
problem.

\begin{table}[H]
\noindent \begin{centering}
\begin{tabular}{|c|c|c|c|c|c|c|c|}
\hline
\multicolumn{8}{|c|}{AR(2) -- Background Error}\tabularnewline
\hline
\hline
\multirow{2}{*}{} & \multirow{2}{*}{$\alpha^{-1}$} & \multicolumn{2}{c|}{${\rm MSE}_{r}$} & \multicolumn{2}{c|}{${\rm MAE}_{r}$} & \multicolumn{2}{c|}{${\rm BIAS}_{r}$}\tabularnewline
\cline{3-8}
 &  & R4D-Var & 4D-Var & R4D-Var & 4D-Var & R4D-Var & 4D-Var\tabularnewline
\hline
\multirow{4}{*}{\textbf{FTH}} & 1 & 0.0254 & 0.0754 & 0.0162 & 0.0629 & 0.0023 & 0.0016\tabularnewline
\cline{2-8}
 & 5 & 0.0328 & 0.0643 & 0.0212 & 0.0534 & 0.0043 & 0.0018\tabularnewline
\cline{2-8}
 & 25 & 0.0722 & - & 0.0608 & - & 0.0187 & -\tabularnewline
\cline{2-8}
 & 50 & 0.0742 & - & 0.0582 & - & 0.0268 & -\tabularnewline
\hline
\multirow{4}{*}{\textbf{WS}} & 1 & 0.0363 & 0.0887 & 0.0272 & 0.0715 & 0.0029 & 0.0012\tabularnewline
\cline{2-8}
 & 5 & 0.0708 & 0.0906 & 0.0571 & 0.0529 & 0.0106 & 0.0017\tabularnewline
\cline{2-8}
 & 25 & 0.0877 & - & 0.0710 & - & 0.0243 & -\tabularnewline
\cline{2-8}
 & 50 & 0.0898 & - & 0.0747 & - & 0.0361 & -\tabularnewline
\hline
\end{tabular}
\par\end{centering}

\caption{Expected values of the ${\rm MSE}_{r}$, ${\rm MAE}_{r}$, and ${\rm BIAS}_{r}$
defined in (\ref{eq:19}), for 30 independent runs. The background
and observation errors are modeled by the first order auto-regressive
($\mathbf{B}=\sigma_{b}^{2}\mathbf{C}_{B}$) and white ($\mathbf{R}=\sigma_{r}^{2}\mathbf{I}$
) Gaussian processes, where $\sigma_{b}=0.10$ (${\rm SNR}\cong10.5$
dB) and $\sigma_{r}=0.08$ (${\rm SNR}\cong12$ dB). The parameter
$\alpha$ denotes the correlation decay rate in the AR(2) covariance
function $\rho(\tau)\propto e^{-\alpha\left|\tau\right|}\left(1+\alpha\left|\tau\right|\right)$.
The studied initial conditions are: flat top-hat (FTH), and window
sinusoid (WS) and the results are reported for both the classic 4D-Var
and the regularized 4D-Var (R4D-Var). The dash lines in the table
denote that the classic method failed to return a solution without
any pre-conditioning. \label{Table2}}
\end{table}

\subsubsection{Selection of the regularization parameters}

\begin{figure}[H]
\noindent \begin{centering}
\includegraphics[scale=0.6]{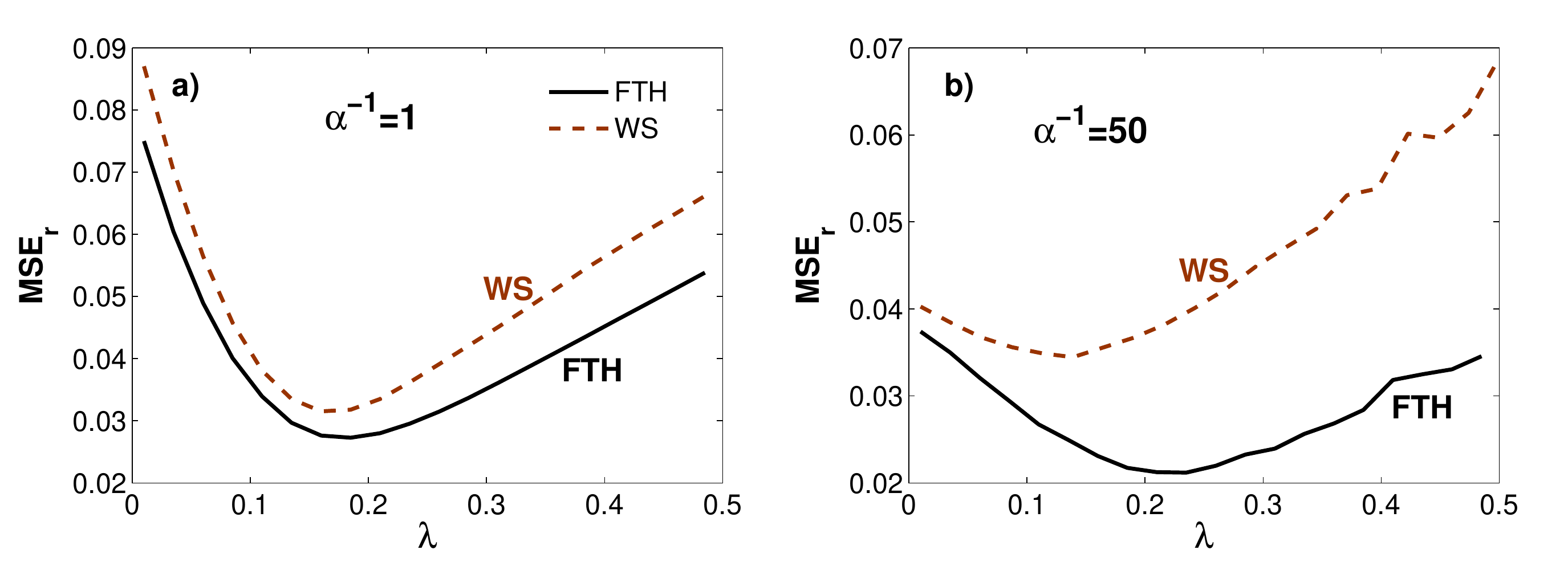}
\par\end{centering}

\caption{The relative mean squared error versus the regularization parameter
obtained for the AR(1) background error for different characteristic
correlation length (a) $\alpha^{-1}=1$, and (b) $\alpha^{-1}=50$.
FTH and WS denote the flat top-hat and window sinusoid initial conditions,
respectively. \label{Fig:8}}

\end{figure}

As previously explained, the regularization parameter $\lambda$ plays
a very important role in making the analysis sufficiently faithful
to the observations and background state, while preserving the underlying
regularity of the analysis. To the best of our knowledge, no general
methodology exists which will produce an exact and closed form solution
for the selection of this parameter, especially for the proposed $\ell_{1}$-norm
regularization \citep[see,][chap.5]{[Han10]}. Here, we chose the
regularization parameter $\lambda$ by trial and error based on a
minimum mean squared error criterion (Figure \ref{Fig:8}). As a rule
of thumb, we found that in general $\lambda\lesssim0.05\left\Vert \mathbf{b}\right\Vert _{\infty}$
yields reasonable results. We also realized that under similar error
signal-to-noise ratio, the selection of $\lambda$ depends on some
important factors such as, the pre-selected basis, the degree of ill-conditioning
of the problem, and more importantly the ratio between the dominant
frequency components of the state and the error.

\section{Summary and Discussion}

We have discussed the concept of sparse regularization in variational
data assimilation and examined a simple but important application
of the proposed problem formulation to the advection-diffusion equation,
relevant to land surface heat and mass flux studies. In particular,
we extended the classic formulations by leveraging sparsity for solving
data assimilation problems in wavelet and spectral domains. The basic
claim is that if the underlying state of interest exhibits sparsity
in a pre-selected basis, this prior information can serve to further
constrain and improve the quality of the analysis cycle and thus the
forecast skill. We demonstrated that the regularized variational data
assimilation (RVDA) not only shows better interpolation properties
but also exhibits improved filtering attributes by effectively removing
small scale noisy features that possibly do not satisfy the underlying
governing physical laws. Furthermore, it is argued that the $\ell_{1}$-norm
RVDA is more robust to the possible ill-conditioning of the data assimilation
problem and leads to more stable analysis compared to the classic
methods.

We explained that, from the statistical point of view, this prior
knowledge speaks for the spatial intrinsic non-Gaussian structure
of the state variable of interest which can be well parameterized
and modeled in a properly chosen basis. We discussed that selection
of the sparsifying basis can be seen as a statistical model selection
problem which can be guided by studying the distribution of the representation
coefficients.

Further research needs to be devoted to developing methodologies to:
(a) characterize the analysis covariance, especially using ensemble
based approaches; (b) automatize the selection of the regularization
parameter and study its impact on various applications of data assimilation
problems; (c) apply the methodology in an incremental setting to tackle
non-linear observation operators \citep{[CouTH94]}; and (d) study
the role of preconditioning on the background error covariance for
very ill-conditioned data assimilation problems in regularized variational
data assimilation settings.

Furthermore, a promising area of future research is that of developing
and testing $\ell_{1}$-norm RVDA to tackle non-linear measurement
and model equations in a hybrid variational-ensemble data setting.
Basically, a crude framework can be cast as follows: (1) given the
analysis and its covariance at previous time step, properly generate
an ensemble of analysis state; (2) use the analysis ensembles to generate
forecasts or background ensembles via the model equation and then
compute the background ensemble mean and covariance; (3) given the
background ensembles, obtain observation ensembles via the observation
equation and then obtain the ensemble observation covariance; (4)
solve an $\ell_{1}$-norm RVDA problem similar to that of (\ref{eq:12})
for each ensemble to obtain ensemble analysis states at present time;
(5) compute the ensemble analysis mean and covariance and use them
to forecast the next time step; and (6) repeat the recursion.

\section*{Acknowledgment}

This work has been mainly supported by a NASA Earth and Space Science
Fellowship (NESSF-NNX12AN45H), a Doctoral Dissertation Fellowship
(DDF) of the University of Minnesota Graduate School to the first
author, and the NASA Global Precipitation Measurement award (NNX07AD33G).
Partial support by an NSF award (DMS-09-56072) to the third author
is also greatly acknowledged.

\appendix
\numberwithin{equation}{section}

\section{Appendix}

\subsection{Quadratic Programming form of the $\ell_{1}$-norm RVDA}

To obtain the quadratic programming (QP) form presented in (\ref{eq:13}),
we follow the general strategy proposed in the seminal work by \citet{[CheDS01]}.
To this end, let us expand the $\ell_{1}$-norm regularized variational
data assimilation ($\ell_{1}$-RVDA) problem in (\ref{eq:12}) as
follows:
\begin{equation}
\underset{\mathbf{x}_{0}}{{\rm minimize}}\,\,\,\left\{ \frac{1}{2}\mathbf{x}_{0}^{{\rm T}}\left(\mathbf{B}^{-1}+\underline{\mathbf{H}}{}^{{\rm T}}\mathbf{R}^{-1}\mathbf{\underline{\mathbf{H}}}\right)\mathbf{x}_{0}-\left(\mathbf{B}^{-1}\mathbf{x}_{0}^{b}+\mathbf{\underline{\mathbf{H}}}^{{\rm T}}\mathbf{R}^{-1}\underline{\mathbf{y}}\right)^{{\rm T}}\mathbf{x}_{0}+\lambda\left\Vert \mathbf{\Phi}\mathbf{x}_{0}\right\Vert _{1}\right\} .\label{eq:A1}
\end{equation}

Assuming $\mathbf{c}_{0}=\mathbf{\Phi}\mathbf{x}_{0}\in\mathbb{R}^{m}$,
then the above problem can be rewritten as,
\begin{equation}
\underset{\mathbf{z}_{0}}{{\rm minimize}}\,\,\,\left\{ \frac{1}{2}\mathbf{c}_{0}^{{\rm T}}\mathbf{Q}\mathbf{c}_{0}+\mathbf{b}^{{\rm T}}\mathbf{c}_{0}+\lambda\left\Vert \mathbf{c}_{0}\right\Vert _{1}\right\} ,\label{eq:A2}
\end{equation}

where, $\mathbf{Q}=\mathbf{\Phi}^{-{\rm T}}\left(\mathbf{B}^{-1}+\mathbf{\underline{\mathbf{H}}}^{{\rm T}}\mathbf{R}^{-1}\underline{\mathbf{H}}\right)\mathbf{\Phi}^{-1}$
and $\mathbf{b}=-\mathbf{\Phi}^{-{\rm T}}\left(\mathbf{B}^{-1}\mathbf{x}_{0}^{b}+\underline{\mathbf{H}}^{{\rm T}}\mathbf{R}^{-1}\underline{\mathbf{y}}\right)$.
Having $\mathbf{c}_{0}=\mathbf{u}_{0}-\mathbf{v}_{0}$, where $\mathbf{u}_{0}=\max\left(\mathbf{c}_{0},\,0\right)\in\mathbb{R}^{m}$
and $\mathbf{v}_{0}=\max\left(-\mathbf{c}_{0},\,0\right)\in\mathbb{R}^{m}$
encode the positive and negative components of $\mathbf{c}_{0}$,
problem (\ref{eq:A2}) can be represented as follows:
\begin{align}
\underset{\mathbf{x}_{0}}{{\rm minimize}\,\,\,} & \left\{ \frac{1}{2}\left(\mathbf{u}_{0}-\mathbf{v}_{0}\right)^{{\rm T}}\mathbf{Q}\left(\mathbf{u}_{0}-\mathbf{v}_{0}\right)+\mathbf{b}^{{\rm T}}\left(\mathbf{u}_{0}-\mathbf{v}_{0}\right)+\lambda1_{m}^{{\rm T}}\left(\mathbf{u}_{0}+\mathbf{v}_{0}\right)\right\} \nonumber \\
 & \,\,\,\,\,\,\,\,\,\,{\rm subject}\,{\rm to\,\,\,\,\,\,}\mathbf{u}_{0}\succcurlyeq0,\,\mathbf{v}_{0}\succcurlyeq0\label{eq:A3}
\end{align}
Stacking $\mathbf{u}_{0}$ and $\mathbf{v}_{0}$ in $\mathbf{w}_{0}=[\mathbf{u}_{0}^{{\rm T}},\,\mathbf{v}_{0}^{{\rm T}}]^{{\rm T}}$,
the more standard QP formulation of the problem is immediately followed
as:
\begin{align}
\underset{\mathbf{w}_{0}}{{\rm minimize}}\,\,\, & \left\{ \frac{1}{2}\mathbf{w}_{0}^{{\rm T}}\begin{bmatrix}\begin{array}{rr}
\mathbf{Q} & -\mathbf{Q}\\
-\mathbf{Q} & \mathbf{Q}
\end{array}\end{bmatrix}\mathbf{w}_{0}+\left(\lambda\mathbf{1}_{2m}+\begin{bmatrix}\begin{array}{r}
\mathbf{b}\\
-\mathbf{b}
\end{array}\end{bmatrix}\right)^{{\rm T}}\mathbf{w}_{0}\right\} \nonumber \\
 & \,\,\,\,\,\,\,\,\,\,{\rm subject}\,{\rm to\,\,\,\,\,\,}\mathbf{w}_{0}\succcurlyeq0.\label{eq:A4}
\end{align}
Obtaining $\mathbf{\hat{w}}_{0}=[\mathbf{\hat{u}}_{0}^{{\rm T}},\,\mathbf{\hat{v}}_{0}^{{\rm T}}]^{{\rm T}}\in\mathbb{R}^{2m}$
as the solution of (\ref{eq:A4}), one can easily recover $\mathbf{\hat{c}}_{0}=\hat{\mathbf{u}}_{0}-\hat{\mathbf{v}}_{0}$
and thus the initial state of interest $\mathbf{\hat{x}}_{0}=\mathbf{\Phi}^{-1}\hat{\mathbf{c}}_{0}$.

The dimension of the QP representation (\ref{eq:A4}) is twice that
of the original $\ell_{1}$-RVDA problem (\ref{eq:A1}). However,
using iterative first order gradient based methods, which are often
the only practical option for large-scale data assimilation problems,
it is easy to show that the effect of this dimensionality enlargement
is minor on the overall cost of the problem. Because, one can easily
see that obtaining the gradient of the cost function in (\ref{eq:A4})
only requires to compute
\[
\begin{bmatrix}\begin{array}{rr}
\mathbf{Q} & -\mathbf{Q}\\
-\mathbf{Q} & \mathbf{Q}
\end{array}\end{bmatrix}\mathbf{w}_{0}=\begin{bmatrix}\begin{array}{r}
\mathbf{Q}\left(\mathbf{u}_{0}-\mathbf{v}_{0}\right)\\
-\mathbf{Q}\left(\mathbf{u}_{0}-\mathbf{v}_{0}\right)
\end{array}\end{bmatrix},
\]
which mainly requires matrix-vector multiplication in $\mathbb{R}^{m}$
\citep[see; e.g.,][]{[FigNW07]}.

\subsection{Upper Bound of the Regularization Parameter}

Here to derive the upper bound for the regularization parameter in
the $\ell_{1}$-RVDA problem, we follow a similar approach as suggested
for example by \citet{[KimKLB07]}. Let us refer back to the problem
(\ref{eq:A2}) which is convex but not differentiable at the origin.
Obviously, $\mathbf{c}_{0}^{a}$ is a minimizer if and only if the
cost function $\mathcal{J}_{R4D}(\mathbf{c}_{0})$ in (\ref{eq:A2})
is sub-differentiable at $\mathbf{c}_{0}^{a}$ and thus
\[
0\in\partial\mathcal{J}_{R4D}(\mathbf{c}_{0}^{a}),
\]
where, $\partial\mathcal{J}_{R4D}(\mathbf{c}_{0}^{a})$ denotes the
sub-differential set at the solution point or analysis coefficients
in the selected basis. Given that
\[
\partial\mathcal{J}_{R4D}(\mathbf{c}_{0}^{a})=\mathbf{Q}\mathbf{c}_{0}^{a}+\mathbf{b}+\lambda\partial\left(\left\Vert \mathbf{c}_{0}^{a}\right\Vert _{1}\right),
\]
we have
\[
-\mathbf{Q}\mathbf{c}_{0}^{a}-\mathbf{b}\in\lambda\partial\left(\left\Vert \mathbf{c}_{0}^{a}\right\Vert _{1}\right).
\]
and thus for $\mathbf{c}_{0}^{a}=\mathbf{0}_{m}$, $\mathbf{0}_{m}=\left[0,\,\ldots,\,0\right]^{{\rm T}}\in\mathbb{R}^{m}$,
one can obtain the following vector inequality
\[
-\lambda\mathbf{1}_{m}\preceq-\mathbf{b}\preceq\lambda\mathbf{1}_{m},
\]
which implies that $\left\Vert \mathbf{b}\right\Vert _{\infty}\leq\lambda$.
Therefore $\lambda$ must be less than $\left\Vert \mathbf{b}\right\Vert _{\infty}$
to obtain nonzero analysis coefficients in problem (\ref{eq:A2})
and thus (\ref{eq:A1}).

\subsection{Gradient Projection Method }

Gradient projection (GP) method is an efficient and convergent optimization
method to solve convex optimization problems over convex sets \citep[see,][pp. 228]{[Ber99]}.
This method is of particular interest, especially, when the constraints
form a convex set $\mathcal{C}$ with simple projection operator.
The cost function $\mathcal{J}_{R4D}(\mathbf{w}_{0})$ in (\ref{eq:13})
is a quadratic function that need to be minimized on non-negative
orthant $\mathcal{C}=\{\mathbf{w}_{0}|\,\, w_{0,i}\geq0\,\,\forall\, i=1,\ldots,2m\}$
as follows:
\begin{eqnarray}
\hat{\mathbf{w}}_{0} & = & {\rm argmin}\,\left\{ \mathcal{J}_{R4D}(\mathbf{w}_{0})\right\} \nonumber \\
 &  & \,\,{\rm s.t.\,\,\mathbf{w}_{0}\succeq0}.\label{eq:A5}
\end{eqnarray}
For this particular problem, the GP method amounts obtaining the following
fixed point:
\begin{equation}
\mathbf{w}_{0}^{*}=\left[\mathbf{w}_{0}^{*}-\beta\nabla\mathcal{J}_{R4D}(\mathbf{w}_{0}^{*})\right]^{+},
\end{equation}
where $\beta$ is a stepsize along the descent direction and for every
element of $\mathbf{w}_{0}$
\begin{equation}
\left[w_{0}\right]^{+}=\begin{cases}
0 & \,\,{\rm if}\,\, w_{0}\leq0\\
w_{0} & {\rm otherwise,}
\end{cases}\label{eq:A6}
\end{equation}
denotes the Euclidean projection operator onto the non-negative orthant.
As is evident, the fixed point can be obtained iteratively as
\begin{equation}
\mathbf{w}_{0}^{k+1}=\left[\mathbf{w}_{0}^{k}-\beta^{k}\nabla\mathcal{J}_{R4D}(\mathbf{w}_{0}^{k})\right]^{+}.
\end{equation}
Thus, if the descent at step $k$ is feasible, that is $\mathbf{w}_{0}^{k}-\beta^{k}\nabla\mathcal{J}_{R4D}(\mathbf{w}_{0}^{k})\succeq0$,
the GP iteration becomes an ordinary unconstrained steepest descent
method, otherwise the result is mapped back onto the feasible set
by the projection operator in (\ref{eq:A6}). In effect, the GP method
finds iteratively the closest feasible point in the constraint set
to the solution of the original unconstrained minimization.

In our study, the stepsize $\beta^{k}$ was selected using the\emph{
Armijo rule,} or the so-called\emph{ backtracking line search}, that
is a convergent and very effective stepsize rule. This stepsize rule
depends on two constants $0<\xi<0.5$ , $0<\varsigma<1$ and assumed
to be $\beta^{k}=\varsigma^{m_{k}}$, where $m_{k}$ is the smallest
non-negative integer for which
\begin{equation}
\mathcal{J}_{R4D}\left(\mathbf{w}_{0}^{k}-\beta^{k}\nabla\mathcal{J}_{R4D}(\mathbf{w}_{0}^{k})\right)\leq\mathcal{J}_{R4D}(\mathbf{w}_{0}^{k})-\xi\beta^{k}\nabla\mathcal{J}_{R4D}(\mathbf{w}_{0}^{k})^{{\rm T}}\nabla\mathcal{J}_{R4D}(\mathbf{w}_{0}^{k}).
\end{equation}
In our experiments the backtracking parameters are set to $\xi=0.2$
and $\varsigma=0.5$ \citep[see,][pp.464 for more explanation]{[BoyV04]}.
In our coding, the iterations terminate if $\frac{\left\Vert \mathbf{w}_{0}^{k}-\mathbf{w}_{0}^{k-1}\right\Vert _{2}}{\left\Vert \mathbf{w}_{0}^{k-1}\right\Vert _{2}}\leq10^{-5}$
or the number of iterations exceeds 100.

\end{document}